\documentclass[12pt,english,cite]{article}
\usepackage{epsfig}
\usepackage{psfrag}

\newcommand{\be}{\begin{equation}}
\newcommand{\ee}{\end{equation}} 
\newcommand{\beq}{\begin{eqnarray}}
\newcommand{\eeq}{\end{eqnarray}}

\def\barnue{\hbox{${\bar \nu}_e$ }}

\title{Supernova neutrinos: Earth matter effects and neutrino mass spectrum}

\author{C.Lunardini$^{a)}$, A.Yu.Smirnov$^{b)}$}

\begin{document}

\def \lta {\mathrel{\vcenter{\hbox{$<$}\nointerlineskip\hbox{$\sim$}}}}
\def \gta {\mathrel{\vcenter{\hbox{$>$}\nointerlineskip\hbox{$\sim$}}}}

\bibliographystyle{h-elsevier2}

\maketitle

\noindent
\begin{center}
{\small {\it a) SISSA-ISAS, via Beirut 2-4, 34100 Trieste, Italy} \\
{\it and INFN, sezione di Trieste, via Valerio 2, 34127 Trieste, Italy}\\
\vspace{0.4cm}
{\it b) The Abdus Salam ICTP, Strada Costiera 11, 34100 Trieste,
Italy} \\
{\it and Institute for Nuclear Research, RAS, Moscow, Russia}}
\end{center}
\vspace{0.5cm}

\begin{abstract}
We perform a detailed study of the Earth matter effects on  supernova neutrinos.  
The dependences of these effects on the properties of the original neutrino fluxes, on the trajectory of the neutrinos inside the Earth and on the  oscillation parameters are described.  
We show that, for a large fraction ($\sim 60\%$) of the possible arrival times of the signal, the neutrino flux crosses a substantial amount of the matter of the Earth at least for one of the existing detectors.
For oscillation parameters from the LMA solution of the solar neutrino problem the Earth matter effect consists in an oscillatory modulation of the $\barnue$ and/or $\nu_e$ energy spectra. The relative deviation with respect to the undistorted spectra can be as large as $20-30\%$ for $E\gta 20$ MeV and  
$70-100\%$ for $E\gta 40$ MeV.  
For parameters from the SMA and LOW solutions the effect is localized at low energies ($E\lta 10$ MeV) and is not larger than $\sim 10\%$.  
The Earth matter effects can be revealed  (i) by the observation of oscillatory distortions of the energy spectra in a single experiment and (ii) by the comparison between the spectra at different detectors. For a supernova at distance $D=10$ Kpc, comparing the results of SuperKamiokande (SK), SNO and LVD experiments one can establish the effect at $(2-3)~\sigma$ level, whereas larger statistical significance ($(4-5)~\sigma$) is obtained if  two experiments of SK-size or larger are available.
Studies of the Earth matter effect will select or confirm the solution of the solar neutrino problem, probe the mixing $U_{e 3}$ and identify the hierarchy of the neutrino mass spectrum.       
\end{abstract}

\section{Introduction}
\label{sec:1}
Neutrinos from gravitational collapses of stars stop to oscillate in vacuum long time before they reach the surface of the Earth. For mass squared 
differences 
implied by the experimental results on solar and atmospheric neutrinos, the supernova neutrinos arrive at Earth  in mass 
eigenstates. 
The reason is  either the loss of coherence between mass eigenstates on the way from the star to 
the Earth or the adiabatic conversion from flavour to mass eigenstates
 inside the star \cite{Mikheev:1986if}. 

After a travel of thousands of years, the neutrinos can still be forced to oscillate again: 
 it is enough 
to put on the way of the neutrinos a filter consisting of flavor non-symmetric 
matter, like 
the Moon or the  Earth. 
The neutrinos will oscillate both in the filter and in vacuum after it. 
Indeed,  the eigenstates of the Hamiltonian in matter do not coincide 
with 
the mass eigenstates $\nu_i$, $i = 1, 2, 3$. 
So the latter turn out to be 
mixed and therefore will oscillate: $\nu_i \leftrightarrow \nu_j$. 
As a consequence, oscillations in the flavour content of the flux are realized.

The possibility of oscillations of supernova neutrinos in the matter of 
the Earth 
has been discussed long time ago \cite{Smirnov:1986ij}. 
It was marked that the  effect 
of oscillations can be significant for values of parameters: 
$\Delta m^2 \sim 10^{-6} - 6 \cdot 10^{-5}$ eV$^2$ and  $\sin^2 2\theta > 
2\cdot 10^{-2}$. 
The effect is different for detectors with different trajectories of the 
neutrinos inside 
the Earth, and studying the oscillation effects in these detectors one 
can restore the direction to the supernova \cite{Smirnov:1986ij}. 

The detection of the neutrino burst from SN1987A 
triggered a number of new 
studies of 
oscillations of  supernova neutrinos and, among them, of oscillations 
inside the Earth. 
It was suggested that the difference in the Kamiokande-2 (K2) and IMB spectra of 
events 
could be related to oscillations of $\bar\nu_e$ 
in the matter of the Earth and to the different 
positions of the detectors at the time of arrival of the burst \cite{talkalexei}. 
For this mechanism to work one needs $\Delta m^2 \sim  10^{-5}$ eV$^2$
and large mixing of the electron neutrinos. 

The first two K2 events showed some directionality with respect to the supernova. The interpretation of these events as due to the scattering of  electron neutrinos from the neutronization peak on electrons put strong bounds on the oscillation parameters, 
excluding a large part of the region which could be relevant for the solar 
neutrinos \cite{Arafune:1987ua}. 
In this region one expects the disappearance of the $\nu_e$ flux due to 
conversion $\nu_e \rightarrow \nu_{\mu}, \nu_{\tau}$ inside the star. It was 
marked however that oscillations inside the Earth can  regenerate the 
$\nu_e$ flux, 
so that the bound is absent in the region of the large mixing angle (LMA) 
solution 
of the solar neutrino problem \cite{Arafune:1987cj,Minakata:1987fj} (see also the discussions in \cite{Lagage:1987xu,Notzold:1987vc}).   In this context, it was shown that the existence of the Earth matter 
effect 
depends on the conversion in the high density resonance associated with the large 
$\Delta m^2$ which is responsible for the oscillations of atmospheric 
neutrinos \cite{Rosen:1988yh,Kuo:1988qu,Minakata:1988cn}.
In particular, the adiabatic conversion in the higher resonance suppresses 
the 
Earth matter effects associated with the lower $\Delta m^2$ which governs solar neutrinos.

At the cooling stage, when the fluxes of all neutrino species are produced, 
the conversion inside the star leads to partial or complete permutation of 
the 
$\bar \nu_e$ and  ${\bar \nu}_{\mu}, {\bar \nu}_{\tau}$ spectra \cite{Wolfenstein:1987pj,Smirnov:1994ku}. 
This causes  the appearance of an high energy tail 
in the $\bar \nu_e$ spectrum which contradicts the SN1987A observations \cite{Smirnov:1994ku}
and therefore implies the exclusion of  some range of the large mixing angles. 
However in the range of $\Delta m^2 \sim  10^{-5}$ eV$^2$  the bound is absent due to the Earth 
matter effect. 

Further studies of the Earth matter effect on supernova neutrinos have been performed in 
ref. \cite{Dighe:2000bi} in connection with the  role the supernova neutrinos can
play in the 
reconstruction of the neutrino mass spectrum. 
The three neutrino schemes both with normal and inverted mass hierarchy 
which explain the atmospheric and the solar neutrino data 
were considered. General formulas 
have been derived  for neutrino and antineutrino fluxes at different 
detectors 
in  presence of the Earth matter effect. 
The main qualitative features 
of the effect have been discussed and some examples of modification 
of the spectrum due to the Earth matter effect were given for  schemes 
with  
 LMA and SMA solutions of the solar neutrino problem. 
The role of the detection of the Earth matter effect  in the identification 
of the neutrino mass scheme was clarified. In particular, 
it was shown that the very fact of the detection of the Earth matter effect
in the neutrino and/or antineutrino channels will allow to establish the 
type of the mass hierarchy and to restrict the element $U_{e3}$ of the 
mixing matrix.

In connection with the fact that the LMA gives the best global 
fit of the solar neutrino data, the interpretation of the SN1987A data has 
been 
revisited \cite{Jegerlehner:1996kx,Lunardini:2000sw,Minakata:2000rx,Kachelriess:2000fe}. The Earth matter effects on the antineutrino fluxes in 
the LMA range 
of the oscillation parameters have been studied in details (the two layer 
approximation of the Earth density profile has been used). 
The regions of the oscillation parameters have been found \cite{Lunardini:2000sw} in which  the Earth 
matter 
effects can explain the difference of the K2 and IMB spectra. Such an 
interpretation 
also favors the normal mass hierarchy case or  very small values of 
$U_{e3}$ for the inverted mass hierarchy \cite{Minakata:2000wm,Lunardini:2000sw,Minakata:2000rx}. 

Recently, the Earth matter effects were considered also in ref. \cite{Takahashi:2000it}
where the expected spectra of events at SuperKamiokande (SK) and SNO have been calculated in three neutrino context with a two-layers approximation for the Earth profile.
\\

In this paper we perform a detailed study of the Earth matter 
effect on the supernova neutrino fluxes 
in the three neutrino schemes which explain the solar and atmospheric 
neutrino data.
Using the realistic Earth density profile,   
we study the dependence of the regeneration effect on the oscillation 
parameters, on the trajectory of the neutrinos
inside the Earth and on the properties of the original neutrino fluxes. 
The effects are calculated  for both neutrinos and antineutrinos.  

We consider the possible arrival directions of the neutrino burst at existing detectors
and  estimate the probability to observe the 
Earth matter effect at least in two detectors. 
The possibility to identify the regeneration effect 
by existing and future experiments is discussed.

The paper is organized as follows. In section \ref{sec:2} we discuss the features of the neutrino fluxes originally produced in the star and the possible trajectories of the neutrinos inside the Earth. The general properties of the Earth matter effects are summarized in sect. \ref{sec:3}; a more specific discussion for LMA oscillation parameters is given in sect. \ref{sec:4}.  In sect. \ref{sec:lowsma} we briefly discuss the regeneration effects for neutrino parameters from the LOW and SMA solutions of the solar neutrino problem.  The possibilities of observation of the Earth matter effects are studied in sect. \ref{sec:6}. Discussion and conclusions follow in sections \ref{disc} and \ref{concl}.  


\section{From the star to the Earth: trajectories and spectra}
\label{sec:2}

\subsection{Inside the star}
\label{sec:2.1}
In this section we summarize the features of the neutrino fluxes as they are produced inside the star and the properties of the supernova that will  be used in our calculations. 
\\

\noindent
A supernova is source of fluxes of neutrinos and antineutrinos of all the three flavours, $e,\mu,\tau$. These fluxes, $F^0_\alpha$ and $F^0_{\bar{\alpha}}$ ($\alpha=e,\mu,\tau$), are characterized by the hierarchy of their average energies,
\be
\langle  E_e  \rangle < \langle  E_{\bar{e}}  \rangle 
< \langle  E_{\mu} \rangle~, 
\label{eq1}
\ee 
and by the equality of fluxes of the non-electron neutrinos (which will be denoted as $\nu_x$): 
\be
F^0_{\mu} =   F^0_{\bar{\mu}} = F^0_{\tau} = F^0_{\bar{\tau}} \equiv 
F^0_{x}~.   
\label{eq2}
\ee    

In absence of neutrino mixing,
$\theta=0$, the neutrino flux at Earth is determined by the original flux produced in the star.
If the decrease of the average energy and of the neutrino luminosity with time occurs over time scales larger than the duration of the burst the flux    of the neutrinos of a given flavour, $\nu_\alpha$, can be described by a Fermi-Dirac spectrum as:
\beq
F^0_\alpha(E,T_\alpha,L_\alpha,D) = 
{L_\alpha \over 4\pi D^2 T^4_\alpha F_3} 
{ E^2 \over e^{E/T_{\alpha}}+1}~, 
\label{eq3} 
\eeq
where $E$ is the
energy of the neutrinos, $L_\alpha$ is the total energy released in 
$\nu_\alpha$  and $T_\alpha$ is the temperature of the $\nu_\alpha$ gas
in the neutrinosphere. Here $D$ represents the distance of the supernova
from the Earth; typically $D\sim 10$ Kpc for a galactic supernova. The quantity $F_3$ is given by
$F_3 = 7\pi^4/120\simeq 5.68$.  According to the hierarchy (\ref{eq1}) 
the indicative values $T_e=3.5$ MeV,  $T_{\bar e}=5$ MeV and $T_x=8$ MeV will be taken as reference in our calculations; results will be presented also for other choices of the temperatures and their dependence on the specific values of $T_e$, $T_{\bar e}$ and $T_x$ will be studied. 
We consider equipartition of the energy between the various flavours, so that $L_\alpha\simeq E_B/6$, with $E_B$ the binding energy emitted in the core collapse of the star: $E_B\simeq 3\cdot 10^{53}~{\rm ergs}$. 
\\
 
In presence of neutrino mixing and masses the neutrinos undergo flavour conversion on their way from the production point in the star to the detector at Earth. 
Matter effects dominate the conversion inside the star, where a wide range of matter densities is met. The conversion effects depend on the distribution of matter in the star; the radial profile
\beq
\rho_s(r)=10^{13}~ C  \left({10 ~{\rm Km}\over r} \right)^3~{\rm g\cdot cm^{-3}} ~,
\label{eq4}
\eeq
with $C\simeq 1 - 15$  provides a good description of the matter distribution for $\rho_s \gta 1~{\rm g\cdot cm^{-3}}$  \cite{BBB,Notzold:1987vc,Kuo:1988qu,Janka}. For $\rho_s \lta 1~{\rm g\cdot cm^{-3}}$ the exact shape of the profile depends on the details of the composition of the star.  For the neutrino parameters we will consider the resonant transitions in the star occur at densities larger than $1~{\rm g\cdot cm^{-3}}$, where the profile (\ref{eq4}) applies; the value $C=4$ will be used unless differently stated. 

\subsection{Crossing the Earth}
\label{sec:2.2}
If the neutrinos cross the Earth before detection, regeneration effects can take place due to the interaction with the matter of the Earth, analogously to what is expected for solar neutrinos \cite{Lisi:1997yc}. 
\\

What is the chance that one, two or even three 
detectors situated in different places of the Earth will detect 
the Earth matter effect on supernova neutrinos?

Due to the short duration of the burst, and the spherical symmetry of the Earth, 
for a given detector the trajectories of neutrinos (and therefore the regeneration effect) can be completely described  by the 
nadir angle $\theta_n$ of the supernova with respect to the detector: if $\cos \theta_n>0$ the detector is shielded by the Earth. 
The angle $\theta_n$ depends (i) on the location of the supernova in Galaxy, (ii) on
the time $t$ of the day  at which the burst arrives at Earth and (iii) on the position of the detector itself. 

We first consider a supernova located in the galactic center (declination\footnote{We define $\delta_s$ as the the angle of the star with respect to the equatorial plane of the Earth.} $\delta_s=-28.9^\circ$) and three detectors \cite{Scholberg:2000ps}: LVD~\cite{Fulgione:1999bk}, SNO~\cite{Virtue:2001mz} and SK~\cite{Fukuda:1998fd}. The positions of these detectors on the Earth are given in the Appendix A.

The  fig.~\ref{fig:nadir} a) shows the dependence of $\cos \theta_n$  
on the time $t$ for the three detectors.
We fixed $t=0$ to be the time at which the star lies on the Greenwich meridian.
The horizontal 
line at $\cos \theta_n = 0.83$ corresponds to the trajectory tangential to
the core of the Earth ($ \theta_n =33.2^\circ$), so that trajectories with 
$\cos \theta_n < 0.83$ are in the mantle of the Earth. For 
$\cos \theta_n > 0.83$ the trajectories cross both the mantle and the core. 

From the figures it appears that:

\begin{enumerate}

\item For most of the arrival times the supernova is seen with substantially different nadir angles at the different detectors, so that
 one expects different Earth matter effects observed. 

\item At any time $t$ the neutrino signal arrives at Earth, at least one detector is shielded by the Earth ($\cos \theta_n>0$) and therefore will see the regeneration effect. Earth shielding is verified even for two detectors simultaneously for a large fraction of the times.

\item At any possible arrival time $t$  one of the detectors is not
shielded by the Earth. So that, once the direction to the supernova is known, 
one can identify such a detector and use its data to reconstruct the neutrino energy
spectrum without regeneration effect. 

\item For a substantial fraction of the times for one of the detectors the trajectory crosses the
core of the Earth. 

\end{enumerate}

In fig.~\ref{fig:nadir} b), c), we show similar dependences of 
$\cos \theta_n$ on the time  $t$ for other locations of the star 
in the galactic plane. Notice that in the case b) two detectors
are not
shielded by the Earth for most of the times and no one trajectory crosses the core. 
In contrast, for the position c) all the three  detectors are always
shielded by the Earth.  In general for the detectors we are considering, which are placed in the northern hemisphere,  the Earth shielding is substantial for a supernova located in the southern hemisphere, as it is the case for stars in the region of the galactic center. 
If a supernova event occurs in the northern hemisphere ($\delta_s>0$), corresponding to some peripherical regions of the galactic disk, the Earth coverage of northern detectors will be scarce, or even null in the limit 
$\delta_s=90^\circ$. In this case southern detectors would be more promising for the observation of Earth matter effects.

Clearly the determination of the position of the supernova is important 
for predictions and the experimental identification of the regeneration
effect. The localization of the star can be done either by 
 direct optical observations or by the experimental study of the 
 neutrino scattering on electrons \cite{Beacom:1998fj},  
which has substantial directionality. Triangulation techniques and neutron recoil methods   
have also been discussed \cite{Beacom:1998fj,Apollonio:1999jg}. As we have mentioned in the Introduction, the study of Earth matter effects by high-statistics experiments can allow to reconstruct the direction to the star. 

\section{Neutrino conversion in the star and in the Earth}
\label{sec:3} 
In this section we summarize the general properties of the Earth matter
effect on  supernova neutrinos. 
We will focus on the 3$\nu$-schemes which explain the
atmospheric and the solar neutrino data.

\subsection{Neutrino mass and mixing schemes}
\label{sec:3.1} 
We assume that 
the atmospheric neutrinos have the
dominant mode of oscillations $\nu_{\mu} \leftrightarrow \nu_{\tau}$ 
with parameters \cite{Fukuda:2000np}: 
\be
|m_3^2 - m_2^2| \equiv    \Delta m^2_{atm} = (1.5 - 4) \cdot 10^{-3} 
{\rm eV}^2, ~~~~~~~ \sin^2 2\theta_{\mu \tau} > 0.88.  
\label{atmpar}
\ee
The solar neutrino data  are explained either by vacuum oscillations (VO solution) or  by one of the MSW solutions 
(LMA, SMA or LOW). The latter are based on the resonant
conversion driven by the oscillation parameters 
\be
|m_2^2 - m_1^2| \equiv   \Delta m^2_{\odot} , ~~~~~~\sin^2 2\theta_{\odot} ~.   
\label{sunpar}
\ee 
Moreover, we consider $\Delta m^2_{atm} \gg \Delta m^2_{\odot} $. 

The electron flavor is distributed in the mass eigenstates 
$\nu_1$ and $\nu_2$ with admixtures  $U_{e1} \approx \cos \theta_{\odot} $, 
$U_{e2} \approx \sin \theta_{\odot} $. We will call 
$\nu_1$ and $\nu_2$ the solar pair of states. 

Three features  of the neutrino schemes, which are important for the supernova neutrino
conversion, are still unknown: 

\begin{enumerate}

\item
The admixture $U_{e3}$ of the $\nu_e$ in the third eigenstate.  Only an upper bound on this 
parameter is known from the  CHOOZ and Palo Verde experiments \cite{chooz,Boehm:2000vp}: 
\be
|U_{e3}|^2 \lta 0.02~. 
\label{ue3}
\ee 

\item 
The type of  mass hierarchy. 
In the case of {\it normal} mass hierarchy the solar pair of  states 
is lighter than $\nu_3$: $m_3 > m_2, m_1$. In the case of inverted  
mass hierarchy the states of the solar pair are heavier than  $\nu_3$:
$m_3 < m_2 \approx m_1$.

\item 
The values of the solar parameters $ \Delta m^2_{\odot} , ~\sin^2 2\theta_{\odot} $.  
Different solutions correspond to  
substantially different values of the oscillation parameters (see e.g. \cite{Bahcall:2001hv}). \\

\end{enumerate}

The masses and mixings determine the pattern of level crossings in the star  
\cite{Dighe:2000bi}. There are two resonances (level crossings) in the 
schemes under consideration:

\begin{itemize}

\item
The high density (H) resonance, determined by the parameters 
$\Delta m^2_{atm}$ and $U_{e3}$. The conversion in the region of this resonance is described by the Landau-Zener type probability, $P_H$, of  
 transition between the mass  
eigenstates $\nu_2$ and $\nu_3$.  

\item
The low density (L) resonance with parameters of the solar pair: 
$\Delta m^2_{\odot} $, $\sin^2 2\theta_{\odot} $. We denote as $P_L$ the probability of $\nu_2 \rightarrow \nu_1$ transition associated to this resonance. 

\end{itemize}

Depending on the type of mass hierarchy and on the value of $\theta_{\odot} $, 
the resonances  appear in different channels. There are four possibilities \cite{Dighe:2000bi}: 

\begin{enumerate}

\item normal mass hierarchy and  $\theta_{\odot}  < \pi/4$: 
both the resonances are in the neutrino channel.

\item normal mass hierarchy and $\theta_{\odot}  > \pi/4$: 
the H resonance is in the neutrino channel, whereas the L resonance is
in the antineutrino channel. This possibility  is disfavored by the
present data on solar neutrinos. 

\item inverted mass hierarchy and $\theta_{\odot}  < \pi/4$: 
the H resonance  is in the antineutrino channel, the L resonance is in the 
neutrino channel. 

\item  inverted mass hierarchy and $\theta_{\odot}  > \pi/4$:
both the resonances are in the antineutrino channel. 

\end{enumerate}

These different schemes correspond to different conversion effects both 
 inside the star and in the matter of 
the Earth.   As it was shown in \cite{Dighe:2000bi}, the Earth effects in the $3\nu$ context depend on (i) the type of mass hierarchy, (ii)  the adiabaticity in the high density resonance, 
 which is determined by   $|U_{e3}|$ and by the density profile of the star, (iii) the oscillation parameters in the low resonance which are determined 
by the solution of the solar neutrino problem.

In what follows we will consider the various possibilities in order.
We will take oscillation parameters from one of the regions of the solutions of 
the solar neutrino problem, 
and also assume that the mixing parameter $|U_{e3}|$ is small, so that  
oscillations 
inside the Earth are reduced to $2\nu$ problem. 

\subsection{Antineutrino channels}
\label{subsec:barnu_chan}
Let us  first consider the scheme with normal mass hierarchy. 

As discussed in sect. \ref{sec:3.1}, in this case   
 there is no level crossing 
in the high resonance region in the antineutrino channel, 
so that the antineutrino flux at the detector does not depend on the jump probability $P_H$. 
We get: 
\be
F^D_{\bar e} = F_{\bar e}  +
(F_{\bar e}^0 - F_x^0)
(1 - 2\bar{P}_L) ({\bar P}_{1e} - |U_{e1}|^2)~, 
\label{banti}
\ee
where 
\be
F_{\bar e} \approx F_{\bar e}^0  -
( F_{\bar e}^0 - F_x^0)
[(1 - \bar{P}_L) - (1 - 2 \bar{P}_L) |U_{e1}|^2]~ 
\label{banti1}
\ee
is the $\barnue$ flux arriving at the surface of the Earth 
(without Earth matter effect) and the fluxes $F_\alpha^0$ are defined in eq. (\ref{eq3}).  
Here ${\bar P}_{1e}$ denotes the probability of ${\bar \nu}_1 \rightarrow {\bar \nu}_e$ conversion  inside the Earth and $\bar{P}_L$ is the jump probability in the L resonance.
The forms (\ref{banti})-(\ref{banti1})  are the consequence of the approximate factorization 
of the dynamics and reduction of the three neutrino problem to an
effective two neutrino conversion (see \cite{Dighe:2000bi} for details).
\\

\noindent
Let us consider the relative Earth effect expressed by the ratio:
\be
{\bar R}\equiv {F_{\bar e}^D-F_{\bar e} \over F_{\bar e}}~. 
\label{ratanti}
\ee
From eqs. (\ref{banti})-(\ref{banti1}) we find 
\be
{\bar R}= {\bar r}(1-2 {\bar P}_L) {\bar f}_{reg}~,
\label{rat2anti}
\ee
where ${\bar r}$ is the (``reduced'') flux factor:
\beq
{\bar r} \equiv \frac{ F_{\bar e}^0 - F_x^0 }
{ F_{\bar e}^0 \left[ \bar{P}_L + (1 - 2 \bar{P}_L) |U_{e1}|^2 \right] + F_x^0 \left[ (1 - \bar{P}_L) - (1 - 2 \bar{P}_L) |U_{e1}|^2 \right]}~,
\label{redanti}
\eeq
and ${\bar f}_{reg}$ the regeneration factor:
\be
\bar{f}_{reg} \equiv ({\bar P}_{1e} - |U_{e1}|^2) ~. 
\label{reganti}
\ee
The three factors present in eq. (\ref{rat2anti}) 
describe the initial conditions 
(initial fluxes) and  different stages of the evolution of the antineutrino
state.   
Let us consider the properties of these factors  in order.\\ 

\noindent 
1. The {\bf L-resonance factor}, $(1 - 2\bar{P}_L)$, is close to the
adiabatic value 1 (i.e. ${\bar P}_L \simeq 0$) especially if the resonance is in the neutrino channel \cite{Kachelriess:2001bs}. 
So, in this case  from eqs. (\ref{rat2anti})-(\ref{redanti}) we get the  simplified expressions: 
\beq
&&{\bar R}= {\bar r} {\bar f}_{reg}~, 
\label{rat2barnu} \\
&&{\bar r} = \frac{ F_{\bar e}^0 - F_x^0 }
{F_{\bar e}^0 |U_{e1}|^2  + F_{x}^0 (1 - |U_{e1}|^2) }~.
\label{redbarnu}
\eeq
\\

\noindent 
2. 
The  {\bf flux factor}, eq. (\ref{redbarnu}), 
determines the sign and the size of the effect. Due to the hierarchy of energies, eq. (\ref{eq1}), a critical energy ${\bar E}_c$ exists at which ${\bar r}=0$. Furthermore 
we have ${\bar r}>0$ below the critical energy, $E<{\bar E}_c$, and ${\bar r}<0$ for $E>{\bar E}_c$.  
 For realistic 
temperatures of the neutrino fluxes (see sect. \ref{sec:2.1}) one gets: 
\be
\bar{E}_c = (25 - 28)~ {\rm MeV}~.
\label{cren_a}
\ee  
At $E\gg {\bar E}_c$ the flux factor (\ref{redbarnu}) is dominated by the harder flux $F^0_x$, so that one finds the asymptotic behavior:
\beq
\bar{r}(E \gg {\bar E}_c) =- \frac{1}{(1 - |U_{e1}|^2)} \simeq -\frac{1}{|U_{e2}|^2~}~.  
\label{banti_h}
\eeq
Similarly, at $E \ll {\bar E}_c$ the flux $F^0_{\bar e}$ dominates, giving the limit: 
\beq
\bar{r}(E \ll {\bar E}_c) = \frac{1}{|U_{e1}|^2}~.
\label{banti_l}
\eeq
From eqs.  (\ref{banti_h})-(\ref{banti_l}) it follows that at very high, as well as at very low energies, the relative regeneration effect (\ref{rat2barnu}) becomes independent of the original fluxes. 
\\

\noindent 
3.
The  {\bf regeneration factor}, eq. (\ref{reganti}), 
describes the propagation effect inside the Earth and is analogous to the regeneration  factor which appears for solar neutrinos. 
Notice that  ${\bar f}_{reg}$  corresponds to genuine matter effect: it
is zero in vacuum. 

The dynamics of propagation and properties of the regeneration factor 
(\ref{reganti}) are different for   
oscillation parameters from different solutions 
of the solar neutrino problem.
\\
 
In what follows we perform numerical calculations of the 
Earth regeneration factor using a realistic density profile 
of the Earth \cite{PREM}. We compare these results 
with results  of the two layers approximation in the 
Appendix B.

\subsection{Neutrino channels}
\label{subsec:nu_chan}
If the hierarchy of the neutrino mass spectrum is normal, the H resonance is in the neutrino channel and the $\nu_e$ flux at the detector depends on $P_H$ \cite{Dighe:2000bi}: 
\be
F^D_e \simeq F_e  +
( F_e^0 - F_x^0)  
P_H (1 - 2P_L) (P_{2e} - |U_{e2}|^2)~, 
\label{basic}
\ee
where the $\nu_e $ flux arriving at the surface of the Earth equals:
\be
F_e \simeq F_e^0  - 
( F_e^0 - F_x^0) [1 -  P_H P_L -  P_H (1 - 2P_L) |U_{e2}|^2]~.  
\label{basic1}
\ee 
Here $P_{2e}$ is the probability of the transition $\nu_2 \rightarrow \nu_e$ inside the Earth. 

From eqs. (\ref{basic})-(\ref{basic1}) one finds the relative Earth matter effect, $R\equiv (F^D_e-F_e)/F_e$, and the flux factor, $r$:
\beq
&&R= r P_H (1-2P_L)f_{reg}~,
\label{rat2}\\
&&r = \frac{ F_e^0 - F_x^0 }
{F_e^0 P_H \left[ P_L + (1-2P_L) |U_{e2}|^2 \right] +F_x^0 \left[1 - P_H P_L- P_H (1-2P_L) |U_{e2}|^2 \right]}~.
\label{red}
\eeq

The regeneration factor, $f_{reg}$, is given by:
\be
f_{reg} \equiv (P_{e2} - |U_{e2}|^2)= -(P_{1e} - |U_{e1}|^2) ~.
\label{regf}
\ee

\noindent
Let us comment on the features of the ratio $R$:
\\

\noindent
1.
From eq. (\ref{rat2}) it follows that if the adiabaticity in the high density (H) resonance inside the star is fulfilled,   
 $P_H \rightarrow 0$,  
the Earth matter effect disappears. The reason  is 
that in the adiabatic case the original electron neutrinos 
convert almost completely into  $\nu_{\mu}$ and $\nu_{\tau}$ 
fluxes in the H resonance. Then the electron neutrinos 
detected at  Earth result from the conversion of the original 
$\nu_{\mu}$ and $\nu_{\tau}$ fluxes. Since these fluxes are 
equal, eq. (\ref{eq2}), no oscillation effect will be observed due to conversion in the 
low density  resonance.  

The Earth matter effect is maximal in  the limit of 
strong violation of the adiabaticity in the H-resonance: 
$P_H \rightarrow 1$, when the dynamics is reduced to a two neutrino
problem with oscillation parameters of the L resonance.

The jump probability $P_H$ is determined by the density profile 
of the star and the oscillation parameters 
$|U_{e3}|^2\approx  \tan^2 \theta_{13}$ and 
$\Delta m^2_{atm}$. In fig.~\ref{fig:ph} we show the lines of equal 
$P_H$ in the $(\Delta m^2_{atm} - \tan^2 \theta_{13})-$ plane,
together with the exclusion region from the CHOOZ experiment.
We use the density profile (\ref{eq4}); the error in $P_H$  due to the uncertainty 
in  the density profile is estimated to be within a factor of 2 \cite{Dighe:2000bi}. 
The figure shows that as $|U_{e3}|^2$ decreases in the range allowed by the bound (\ref{ue3}) the transition in the H resonance varies from perfectly adiabatic ($P_H\simeq 0$), for $|U_{e3}|^2\gta 5\cdot 10^{-4}$, to strongly non-adiabatic  ($P_H\simeq 1$), for $|U_{e3}|^2\lta 10^{-6}$. The intervals of adiabaticity and strong adiabaticity violation change only mildly as $\Delta m^2_{atm} $ varies in the presently allowed range. 
Notice that future atmospheric neutrino studies and the long base-line 
experiments will sharpen the allowed region of 
$\Delta m^2_{atm}$. 
\\

\noindent
2.
The low density resonance factor,
$(1 - 2 P_L)$, 
 is zero if $P_L = 1/2$, which corresponds to 
a situation when the neutrino beam arriving at Earth consists in
incoherent and equal fluxes of $\nu_1$ and $\nu_2$. In this case 
the effects of $\nu_1 \rightarrow \nu_e$  
and $\nu_2 \rightarrow \nu_e$ oscillations cancel each other. 

In fig.~\ref{fig:adiablow} we show the lines of  $P_L = 1/2$, calculated with the density profile (\ref{eq4})  in the 
($\tan^2 \theta_{\odot}  -  \Delta m^2_{\odot} $) plane for different 
values of the neutrino energy.
The lines cross the allowed region 
of the SMA solutions for the energy interval 
$E = 5 - 15$ MeV. 
In fig.~\ref{fig:adiablow} we show also the lines $P_L = 0.05$ for 
two  different energies: $E = 5$ MeV and $E = 50$ MeV. 
These lines determine the lower edge of the 
adiabaticity region in the 
$(\tan^2 \theta_{\odot}  -  \Delta m^2_{\odot} )$ plane. 
Notice that the LMA region is in the adiabaticity
domain for all the relevant energies, whereas the SMA solution region 
is in the domain of significant adiabaticity violation. Depending on the details of the density profile of the star the LOW solution lies either in the adiabaticity region or in the region of partial adiabaticity breaking. 

A qualitative  treatment does not depend on whether the 
low density resonance is in
the neutrino or antineutrino channel (dark side of the parameter space). 
Quantitatively the results are different. 
\\

\noindent
3.
The  flux factor, $r$, eq. (\ref{red}), changes sign at lower 
critical energy with respect to the case of antineutrinos, since the original $ {\nu}_e$ spectrum is softer than the $\barnue$ spectrum. We get: 
\be
E_c = (16 - 24)~ {\rm MeV}~.
\label{cren}
\ee  
Similarly to what was discussed for $\barnue$, the flux factor becomes independent of the original fluxes in the low and high energy limits:
\beq
&&r(E\gg E_c) = -\frac{1 }
{1 - P_H P_L- P_H (1-2P_L) |U_{e2}|^2}~,
\label{redeinf} \\
&&r(E \ll E_c) =
 \frac{1 }
{P_H \left[ P_L + (1-2P_L) |U_{e2}|^2 \right]}~.
\label{redelow}
\eeq
\\

\noindent
4.
The Earth regeneration factor, eq. (\ref{regf}),  
depends on the mixing and mass squared difference of the solar pair and on the nadir angle $\theta_n$. It will be described in detail in the following sections.

\subsection{Schemes with inverted mass hierarchy}
\label{subsec:invhier}
If the hierarchy of the mass spectrum is inverted the high density resonance is in the antineutrino channel (see sect. \ref{sec:3.1}) and the Earth matter effect for ${\bar \nu}_e$ depends on the jump probability $P_H$.
The expressions (\ref{banti})-(\ref{banti1}) for the $\barnue$ fluxes  are immediately generalized to:
\beq
&&F^D_{\bar e} = F_{\bar e}  +
(F_{\bar e}^0 - F_x^0) P_H 
(1 - 2\bar{P}_L) ({\bar P}_{1e} - |U_{e1}|^2)~,
\label{banti1inv}  \\
&&F_{\bar e} \approx F_{\bar e}^0  -
( F_{\bar e}^0 - F_x^0)
[1 - P_H \bar{P}_L - P_H (1 - 2 \bar{P}_L) |U_{e1}|^2]~ ,
\label{banti2inv}
\eeq 
in analogy with eqs. (\ref{basic})-(\ref{basic1}).

Taking $\bar{P}_L=0$ (see sect. \ref{subsec:barnu_chan}), we find the relative deviation, $\bar R$, and the reduced flux factor:
\beq
&&{\bar R}= {\bar r} P_H {\bar f}_{reg}~, 
\label{rat2barnuinv} \\
&&{\bar r} = \frac{ F_{\bar e}^0 - F_x^0 }
{F_{\bar e}^0 P_H |U_{e1}|^2  + F_x^0 (1 - P_H |U_{e1}|^2)  }~,
\label{redbarnuinv}
\eeq
with the asymptotic limits:
\beq
&&{\bar r}(E\gg {\bar E}_c) = -\frac{1 }
{1- P_H |U_{e1}|^2}~,
\label{redantiinf} \\
&&{\bar r}(E \ll {\bar E}_c) = \frac{1 }
{P_H |U_{e1}|^2}~.
\label{redantilow}
\eeq

Clearly, the conversion of $\nu_e$ is independent of $P_H$. The expressions of the neutrino fluxes $F_e^D$ and $F_e$ can be obtained from  eqs. (\ref{basic})-(\ref{basic1}) by the replacement $P_H \rightarrow 1$; they become analogous to eqs. (\ref{banti})-(\ref{banti1}).
With the same prescription, from eqs. (\ref{rat2})-(\ref{red}) one gets the expressions of the ratios $R$ and $r$. 
\\

\noindent
Summarizing the  results of sections   \ref{subsec:barnu_chan}-\ref{subsec:invhier} we can say that the mass hierarchy and  the adiabaticity in the H density resonance (and thus $U_{e3}$) determine the channel ($\nu_e$ or ${\bar \nu}_e$) in which the Earth matter effects appear, which is:

\begin{itemize}

\item both the $\nu_e$ and ${\bar \nu}_e$ channels if the H resonance is strongly non-adiabatic, $P_H=1$, regardless to the hierarchy.

\item the ${\bar \nu}_e$ channel for adiabatic  H resonance,  $P_H=0$, and normal hierarchy.

\item the $\nu_e$ channel for adiabatic  H resonance,  $P_H=0$, and  inverted hierarchy.

\end{itemize}

The possibility of probing $U_{e3}$ and the mass hierarchy by the study of Earth effects on supernova neutrinos will be discussed in sect. \ref{disc}.

\section{The Earth matter effects for the LMA parameters}
\label{sec:4}

\subsection{Antineutrino channels}
\label{sec:4.2}
Let us now study the features of the relative Earth matter effect, eq. (\ref{rat2barnu}), for $\barnue$ with mixing and mass squared difference in the LMA region. 
\\

\noindent
The dynamics of the conversion inside the Earth is described by the regeneration factor ${\bar f}_{reg}$, eq. (\ref{reganti}).
For LMA parameters the Earth matter effect consists in an oscillatory modulation of the neutrino energy spectrum.

The fig. \ref{fig:fig5} shows the ratio ${\bar R}$ as a function of the neutrino energy for various values of $\theta_n$. 
For mantle crossing trajectories, $\theta_n>33.2^\circ$, the effect is mainly due to the interplay of 
oscillations and adiabatic 
evolution. That is, oscillations in medium with slowly varying
density. Small density jumps produce only rather weak  effects. As a result, the features of the regeneration factor, ${\bar f}_{reg}$, are similar to what is predicted in the case of propagation in medium with constant density (see Appendix B).  The factor is positive in the whole energy spectrum, so that the sign of the matter effect is determined by the flux factor (\ref{redbarnu}): we have ${\bar R}>0$ for $E<{\bar E}_c$ and ${\bar R}<0$ for $E>{\bar E}_c$.  

The energy spectrum shows regular oscillations in energy with period
\begin{equation}
\Delta E \approx \frac{2\pi}{\phi (\theta_n, E)} E~,  
\label{period}
\end{equation}
where the oscillation phase $\phi$ is determined by the integral  
\be 
\phi(\theta_n) = 2\pi \int_0^{r(\theta_n)} \frac{dx}{l_m(n(x), E)} 
\label{phase}
\ee
over the  neutrino trajectory. Here $l_m$ is the (instantaneous) 
oscillation length in matter, and $n(x)$ is the electron density along the trajectory. 
As follows from eqs. (\ref{period})-(\ref{phase}) the 
period of  oscillations decreases with the nadir angle 
and increases with the energy. 
The dependence on the energy appears in $\Delta E$, 
(see eq.  (\ref{period})) 
explicitly, and implicitly via the oscillation length. 

As a result of adiabatic evolution,  
the depth of oscillations of the regeneration
factor is determined by the electron number density  at the surface of the
Earth (see Appendix B), $n_e^0$: 
\be
{\bar D}_f \approx 2 \sqrt{2} G_F n_e^0 \frac{E}{\Delta m^2_{\odot}} \sin^2 2\theta_m^0~.
\label{depth}
\ee
Here $\theta_m^0$ is the mixing angle of the solar pair in matter at the
surface.  

The depth ${\bar D}_f$ has a resonant dependence on the quantity $x\equiv 2E |V|/\Delta m^2_{\odot}$, with $V$ being the matter potential (see the Appendix B for details). 
Both ${\bar D}_f$ and $l_m$ increase as the system approaches the resonance; correspondingly, the period $\Delta E/E$ increases. For neutrinos propagating in the mantle and $\Delta m^2_{\odot} = 5 \cdot 10^{-5}$ eV$^2$ (which is used in the figure \ref{fig:fig5}) the resonance 
is realized at  $E=E_R \simeq 150$ MeV. Thus the Earth effect is larger in the highest energy part of the  spectrum. 

For core crossing trajectories the behavior of the Earth effect becomes 
irregular  due to the 
interference of the oscillations  in the core and in the mantle. 
The modulations in the energy spectra have smaller period 
both due to presence of large densities and larger  length of the trajectory. 
Moreover, now the effect can change the sign both below 
and above the critical energy. 
That is, for some energies the Earth effect is negative at low energies
and positive at high energies.

As $\Delta m^2_{\odot}$ decreases, as shown in fig.~\ref{fig:fig7}, the regeneration factor increases, since the resonance of the system is realized at lower energies. In particular, for $\Delta m^2_{\odot} = 2 \cdot 10^{-5}$ eV$^2$, the resonance energy 
equals  $E_R \simeq 60$ MeV.  
The period (in the energy scale) of the oscillatory  modulation 
of the spectrum increases with the decrease of $\Delta m^2_{\odot}$. 

The change of the mixing parameter $|U_{e1}|^2 \approx \cos^2 \theta_{\odot} $
(within the LMA region) 
influences the regeneration factor rather weakly. 
However variations of  $|U_{e1}|^2$ change the spectrum 
arriving at the surface of the Earth. 
According to the eq. (\ref{banti1}), with the increase of 
$|U_{e1}|^2$ (i.e. decrease of the mixing $\theta_{\odot} $), the contribution of the hard component 
in the spectrum decreases: the composite spectrum becomes softer.  
\\

\noindent
Given the mixing $U_{e1}$  the  factor  $\bar r$, eq. (\ref{redbarnu}), depends only on the original fluxes, $F_{\bar e}^0$ and $F_x^0$.  The fig. \ref{fig:fig8} shows the relative Earth effect, $\bar R$, as a function of the energy for different values of $\bar r$ determined by different temperatures of the original $\barnue$ spectrum.  
As it appears in the figure, the critical energy ${\bar E}_c$ decreases with $T_{\bar e}$, so that the region in which the flux factor suppresses the regeneration effect shifts to lower energies. For $E\ll {\bar E}_c$ and $E\gg {\bar E}_c$ the depth of oscillations depends only very weakly on variations of $T_{\bar e}$, 
according to the limits (\ref{banti_l}) and (\ref{banti_h}), which do not depend on temperatures.
The values of $T_{\bar e}$ and $T_x$  only affect the rapidity of the convergence to these limits: the convergence is faster for the larger difference $T_x-T_{\bar e}$.  This appears in fig. \ref{fig:fluxf} b), in which the flux factor $\bar r$ is plotted as a function of the antineutrino energy with the same parameters as in the fig. \ref{fig:fig8}.
For $\sin^2 2\theta_{\odot}=0.75$, as used in the figs. \ref{fig:fig8}-\ref{fig:fluxf}, one finds ${\bar r}(E\ll {\bar E}_c)=1.33$ and ${\bar r}(E\gg {\bar E}_c)=-4$. Thus the Earth effect has stronger enhancement at high energies. 
\\

\noindent 
If the hierarchy is inverted the Earth matter effect on $\barnue$ is affected by the adiabaticity in the high density resonance.
Such dependence is illustrated in fig. \ref{fig:fig6}, which shows the 
the relative effect $\bar R$ as a function of the energy  for various values of $P_H$.  
According to eq. (\ref{rat2barnuinv}) the effect is proportional to $P_H$ and is maximal at $P_H=1$. 
The dependence of $\bar R$ on $P_H$ is transparent in the limits of high and low energies.
Combining eqs. (\ref{rat2barnuinv}) and  (\ref{redantiinf}) we find that at high energies $\bar R$ depends on $P_H$ as:
 \be
{\bar R}(E \gg {\bar E}_c)= -\frac{P_H}
{1- P_H |U_{e1}|^2} {\bar f}_{reg} ~,
\label{Rbarnuinf}
\ee
that is, ${\bar R} \propto P_H$ for $P_H \ll 1$. 
The weak dependence of $\bar R$ on $P_H$ in the softer part of the spectrum in fig. \ref{fig:fig6} is explained by the fact that the dependence on $P_H$ cancels in the low energy limit, $E\ll {\bar E}_c$, as can be seen from eqs. (\ref{rat2barnuinv}) and  (\ref{redantilow}).

\subsection{Neutrino channels}
\label{sec:4.1}
Let us discuss the properties of the Earth regeneration effect in the $\nu_e$ channel.

As it was shown in fig. \ref{fig:adiablow}, for LMA oscillation parameters the adiabaticity in the L-resonance inside the star is satisfied, so that $P_L=0$ and eqs. (\ref{rat2})-(\ref{red}) reduce to: 
\beq
&&R=r P_H f_{reg}~,
\label{redrat}\\
&&r = \frac{ F_e^0 - F_x^0 }
{ F_e^0 P_H |U_{e2}|^2 + F_x^0 (1 -  P_H |U_{e2}|^2) }~.
\label{redrat2}
\eeq
\\

\noindent 
The regeneration factor $f_{reg}$, eq. (\ref{regf}), and therefore $R$, have similar dependence on $\theta_n$ and $\Delta m^2_{\odot}$ as in the case of antineutrinos. These dependences are illustrated in the figs. \ref{fig:fig1} and \ref{fig:fig3}, where $P_H=1$ was taken.
The oscillation length and the period of the modulations in the energy spectrum increase with the increase of the energy and the decrease of $\Delta m^2_{\odot}$ 
(fig. \ref{fig:fig3}). The depth of the oscillations of the regeneration factor $f_{reg}$ is larger than for antineutrinos since (if the L resonance is in the neutrino channel) matter enhances the $\nu_e$ mixing and suppresses the mixing of $\barnue$:  
\be 
\sin^2 2\theta_m (\bar{\nu}) < \sin^2 2\theta_{\odot} < \sin^2 2\theta_m
(\nu)~. 
\ee 
The depth of oscillations has a resonant character (see Appendix B), increasing as the resonance energy is approached. According to eq. (\ref{depth}) 
the depth gets larger for smaller $\Delta m^2_{\odot}$ (fig. \ref{fig:fig3}).
\\

\noindent
The dependence of the Earth matter effect on the flux factor, $r$, eq. (\ref{redrat2}), is illustrated in fig. \ref{fig:fig4}, where the ratio $R$ is plotted for different values of the temperature $T_{ e}$, $T_x=8$ MeV and $P_H=1$.  
According to sects. \ref{subsec:barnu_chan} and  \ref{subsec:nu_chan}  the flux factor suppresses the Earth matter effect at energies close to the critical energy, $E_c$, which is slightly lower than for $\barnue$: $E_c\simeq 22$ MeV for $T_{e}=3.5$ MeV.
At high and low energies the asymptotic limits (\ref{redeinf})-(\ref{redelow}) are realized. This is shown in fig. \ref{fig:fluxf} a), with the same values of the temperatures as in fig. \ref{fig:fig4} and $P_H=1$.  From eqs. (\ref{redeinf})-(\ref{redelow}) (with $P_L=0$) and from the fig.  \ref{fig:fluxf} it follows that, in contrast to the case of $\barnue$, the Earth effect has stronger enhancement at low energy: for $\sin^2 2\theta_{\odot}=0.75$ one gets $r(E\ll E_c)=4$ and $r(E\gg E_c)=-1.33$. Notice that the convergence to these limits is faster than for $\barnue$ due to larger difference of the $\nu_e$ and $\nu_x$ temperatures. 
\\

\noindent
According to eq. (\ref{redrat}), the Earth matter effect is larger for larger $P_H$, i.e. for maximal adiabaticity breaking in the high density resonance inside the star.
From eqs. (\ref{redrat}) and (\ref{redrat2}) the following asymptotic limit follows:
\be
R(E \gg E_c)= -\frac{P_H}
{1- P_H |U_{e2}|^2} f_{reg}~,
\label{Rinf}
\ee
similarly to eq. (\ref{Rbarnuinf}). The combination of eqs. (\ref{redrat}) and (\ref{redelow}) implies that $R$ becomes independent of $P_H$ at $E\ll E_c$.  These features of the dependence of $R$ on $P_H$ are shown in fig. \ref{fig:fig2}. 
\\

As discussed in sect. \ref{subsec:invhier}, the results for inverted hierarchy of the spectrum are obtained from the description given for normal hierarchy by the replacement $P_H \rightarrow 1$. Therefore the results shown in the figs. \ref{fig:fig1}-\ref{fig:fig4}, in which $P_H=1$ was used, apply to the case of inverted hierarchy.

\section{The case of  oscillation parameters in the LOW and SMA regions} 
\label{sec:lowsma}

\subsection{LOW parameters} 
\label{subsec:low}
For oscillation parameters in the region of the LOW solution of the solar neutrino problem the mass squared difference is smaller by at least two orders of magnitude with respect to the LMA solution: $\Delta m^2_{\odot} \sim 10^{-8}-10^{-7}~{\rm eV^2}$. Therefore   the resonance energy in the Earth is very small, $E_R< 1$ MeV, and in the whole energy spectrum of supernova neutrinos the neutrino system is far above the resonance. 
Both the $\nu_e$ and $\barnue$ mixings are suppressed in the matter of the Earth and the oscillation lengths approach the refraction length, $l_m \approx l_0\sim 8000$ Km.  As a consequence, the regeneration inside the Earth has only weak dependence on the neutrino energy and no oscillatory distortions appear in the energy spectra at the detectors. In contrast, the regeneration effect  depends strongly on the nadir angle $\theta_n$: for  $E=10$ MeV the regeneration factor has a maximum,  $f_{reg}\simeq 0.04$,  for $\theta_n\simeq 25^\circ$ (see e.g. \cite{Gonzalez-Garcia:2000dj}). 

The fig. \ref{fig:fig10} shows the relative Earth effects for $\nu_e$  and $\barnue$  as functions of the energy for  $\Delta m^2_{\odot}=10^{-7}~{\rm eV^2}$, $\sin^2 2\theta_{\odot}=0.9$, $\theta_n=25^\circ$ and $P_H=1$. We have also considered perfect adiabaticity in the L resonance inside the star, $P_L=0$.
The figure shows that for $E>5$ MeV the effect is not larger than $20\%$ for $\nu_e$ and $10\%$ for $\barnue$; it decreases with the increase of energy with a $\sim 1/E$ behavior.
In the neutrino channel the relative deviation $R$, eq. (\ref{rat2}), is larger than in the antineutrino channel, especially in the very soft part of the spectrum, $E\lta 10$ MeV. This is explained (i) by the fact that the neutrino system approaches the resonance at low energies, and therefore the $\nu_e$ mixing is enhanced, and (ii) by the larger flux factor in the low energy limit, $E\ll E_c$ (see  fig. \ref{fig:fluxf}).
We get $\sim 50\%$ effect at $E=2$ MeV and $\sim 100\%$ at $E=1$ MeV.

\subsection{SMA parameters} 
\label{subsec:sma}
For oscillation parameters in the SMA region the fluxes at the detector,  $F^D_{\bar e}$ and $F^D_e$,  eqs. (\ref{banti}) and (\ref{basic}), are substantially different with respect to the case of LMA parameters, due to differences in the L resonance factor, $(1-2P_L)$, and in the regeneration factor.

In what follows we describe the Earth effect for $\nu_e$; effects on $\barnue$ conversion are extremely small due to the suppression of the mixing in matter and will not be considered.
\\

\noindent
 Let us first discuss the factor $(1-2P_L)$.  For $\Delta m^2_{\odot}$ and $\sin^2 2\theta_{\odot}$ in the SMA region the conversion in the L resonance inside the star occurs in the adiabaticity breaking regime (fig. \ref{fig:adiablow}). The jump probability $P_L$ differs from $0$ significantly and, in particular, as shown in fig.  \ref{fig:adiablow}, $P_L$ is close to $1/2$, thus suppressing the matter effect (eq. (\ref{rat2})). 

In fig. \ref{fig:pl} we show the factor $(1-2P_L)$ as a function of the neutrino energy for $\Delta m^2_{\odot}=6\cdot 10^{-6}~{\rm eV^2}$ and two values of $\sin^2 2\theta_{\odot}$ in the SMA region. We used the profile (\ref{eq4}) with $C=4$. 
For the 
largest possible mixing, $\sin^2 2\theta_{\odot} \sim 5 \cdot 10^{-3}$, the factor is
negative above 10 MeV; for the best fit point, $\sin^2 2\theta_{\odot} \sim 2.4 \cdot 10^{-3}$, it is negative in the whole
detectable part of the spectrum ($E > 5$ MeV).  
These results, however, strongly depend on the model of the star. 
For massive stars ($M > 30 M_{\odot}$)  the density profile may have 
smaller gradient, so that the adiabaticity breaking is 
weaker. In this case the energy at which $P_L = 1/2$ is larger, 
and in a significant  part of the spectrum the low resonance factor 
can be  positive.  
Notice that in the high energy  part of the spectrum  
the factor can be as large  as 0.7 - 0.8 in absolute value. 
\\

\noindent
The regeneration factor, $f_{reg}$, eq. (\ref{regf}), has a peculiar resonant behaviour in dependence on the  energy and on the nadir angle $\theta_n$. 
For mantle crossing trajectories, $\theta_n>33.2^\circ$, the matter effect consists in resonantly enhanced oscillations, with oscillation length comparable or larger than the radius of the Earth.  A peak appears in the $\nu_e$ spectrum at $E\sim 15 $ MeV.

For core crossing trajectories, $\theta_n<33.2^\circ$, the regeneration factor exhibits a narrow peak at $E\simeq 7$ MeV due to parametric resonance, with two smaller peaks at higher and lower energy due to MSW resonances in the mantle and in the core respectively.  These features are shown in the fig. \ref{fig:fig9}, which represents the relative deviation $R$ as a function of the energy for $P_H=1$, $\theta_n=0^\circ$ and various values of the factor $C$ in the density profile (\ref{eq4}), corresponding to different values of $P_L$.
With increase of $C$ the jump probability $P_L$ decreases, 
so that the region where $(1 - 2 P_L) > 0$ expands to higher 
energies. With the decrease of $P_L$ the Earth matter effect changes from negative to positive and the
size of the effect increases; it can be as large as $\sim 30\%$. 

As in the case of the LMA solution, the effect decreases with $P_H$  and disappears for perfectly adiabatic transition in the H resonance, $P_H=0$. 
If the mass hierarchy is inverted the matter effect exists, in the neutrino channel,  independently of the character of the H resonance. 

\section{Observation of Earth effects}
\label{sec:6}
Let us first summarize the results of the previous sections.

The regeneration effects in the matter of the Earth produce a distortion of the neutrino energy spectrum.  The character of the distortion depends (i) on  the properties of the original neutrino and antineutrino fluxes, (ii) on the nadir angle $\theta_n$ and  (iii) on the features of the neutrino masses and mixings, in particular on the solution of the solar neutrino problem.

For oscillation parameters from the LMA solution the distortion of the spectrum has oscillatory character with larger oscillation depth and period in the high energy part of the spectrum. The effect exists  in the neutrino or in the antineutrino channel or in both depending on the adiabaticity in the H resonance and on the mass hierarchy. 

For the LOW solution there is no oscillatory behaviour in the spectra. The effect is very small in the whole observable part of the energy spectrum and decreases with the increase of the energy. 

For parameters in the region of the SMA solution the Earth effect exists for neutrinos only and consists in the appearance of one or three (for core crossing trajectories) resonance peaks (or dips) in the low energy part of the spectrum. 

In this section we discuss the possibility of detecting the Earth matter effects at present and future detectors.

\subsection{Detection of supernova neutrinos. Numbers of events}
\label{subsec:6.1} 
As follows from the results of the previous sections, the observation of the Earth matter effect requires: (i) separate detection of neutrinos of different flavours, (ii) separate detection of neutrinos and antineutrinos, (iii) the reconstruction of the neutrino energy spectrum.

In what follows we concentrate on events from charged current (CC) scattering on nucleons or nuclei at real-time detectors. These events better satisfy the requirements (i)-(iii); in particular CC processes have high sensitivity to the neutrino energy spectrum. 
\\

We consider:

\begin{enumerate}

\item  The detection of $\barnue$ at water Cerenkov detectors (SuperKamiokande and the outer volume of SNO) via the reaction:
\be
\bar{\nu}_e + p \rightarrow e^+  +  n~.
\label{antinu_r}
\ee
Other CC reactions (e.g. the scattering of $\nu_e$ and $\barnue$ on oxygen nuclei) have substantially smaller cross section so that they contribute to the total number of events at few per cent level; 
they will not be considered further.

\item Heavy water detectors (the inner volume of SNO experiment) with the detection reactions:
\beq
\nu_e + d \rightarrow e  +  p  + p~,
\label{nu_rd} \\
\bar{\nu}_e + d \rightarrow e^+  +  n + n~,
\label{antinu_rd}
\eeq
which represent the dominant channel of CC detection.  Events from the process (\ref{antinu_rd}) will be distinguished by those from (\ref{nu_rd}) if 
 neutrons are efficiently detected 
in correlation with the positron.

\item Liquid scintillator detectors (LVD), which are mostly sensitive to $\barnue$ via the reaction (\ref{antinu_r}) with only little sensitivity to absorption processes on carbon nuclei.

\end{enumerate}

Besides pure CC processes, the reactions  
\beq
&&\nu_i + e \rightarrow \nu_i + e~, ~~~~ i =e, \mu, \tau, \\ 
&&{\bar \nu}_i + e \rightarrow {\bar \nu}_i + e~,
\label{nu_rx}
\eeq 
and the NC breakup of deuterium:
\beq
&&\nu_i + d \rightarrow \nu_i + n+p ~, ~~~~ i =e, \mu, \tau, \\ 
&&{\bar \nu}_i + d \rightarrow {\bar \nu}_i + n+p ~,
\label{NCd}
\eeq 
allow to  reconstruct the total neutrino flux; moreover, due to its good directionality, the scattering of $\nu_e$ on electrons is relevant to the location of the supernova.

Radiochemical experiments could provide information on the total $\nu_e$ flux above a certain threshold. 

The number of CC events with lepton having the observed kinetic energy 
$E_e$ is given by 
\begin{eqnarray}
{dN_{\alpha} \over dE_e} = 
N_T\int_{-\infty}^{+\infty} dE_e' R(E_e,E_e') 
{\cal E}(E_e') \int dE F_{\alpha} (E) 
{\sigma (E_e', E) \over dE_e'}~, 
\label{eq:dnem}
\end{eqnarray}
where  $E_e'$ is the true energy of the electron (or positron), $N_T$ is the number of target particles in the fiducial volume and  ${\cal E}$ represents  the detection efficiency. Here ${\sigma (E_e', E)/dE_e'}$ is the differential cross section and 
$R(E_e, E_e')$ is the energy resolution function, which 
can be described by a gaussian form: 
\begin{eqnarray} 
R(E_e ,E_e') = {1 \over \Delta\sqrt{2 \pi}} 
\exp \left[- { (E_e - E_e')^2 \over 2 \Delta^2} \right]~.
\label{eq:resfun} 
\end{eqnarray}  
The energy resolution $\Delta$ and the other parameters of the detectors (volume, efficiency, etc.) are summarized in the Appendix A.

The energy spectrum (\ref{eq:dnem}) of the charged leptons reflects the spectrum of the neutrinos, with the following differences:

\begin{itemize}

\item the energy dependence of the cross section\footnote{The $\sigma \propto E^2$ dependence constitutes a good approximation at low energies; in the highest energy part of the supernova neutrino spectrum deviations due to weak magnetism and recoil effects are relevant, see e.g. \cite{Vogel:1984hi,Fayans:1985ru,Vogel:1999zy}. In our calculations we used the cross sections in ref. \cite{Vogel:1999zy} for the scattering (\ref{antinu_r}) and in ref. \cite{Nakamura:2000vp} for the reactions (\ref{nu_rd})-(\ref{antinu_rd}). },  $\sigma \propto E^2$, substantially enhances the high energy part of the spectrum. 

\item  the integration over the neutrino energy and the convolution with the energy resolution function, eq. (\ref{eq:dnem}), lead to averaging out the fast modulations in low energy part of the spectrum (appearing for LMA oscillation parameters).
Conversely, the large-period oscillations at high energies will appear in the lepton spectrum (\ref{eq:dnem}).

\end{itemize}

\subsection{Identification  of the  Earth matter effects}
\label{subsec:6.2}
The Earth matter effects can be identified:

\begin{enumerate}

\item at a single detector, by the observation of deviations of the energy spectrum with respect to what expected from conversion in the star only. 

\item by the comparison of energy spectra from different detectors.

\end{enumerate}

\begin{figure}[hbt]
\begin{center}
\epsfig{file=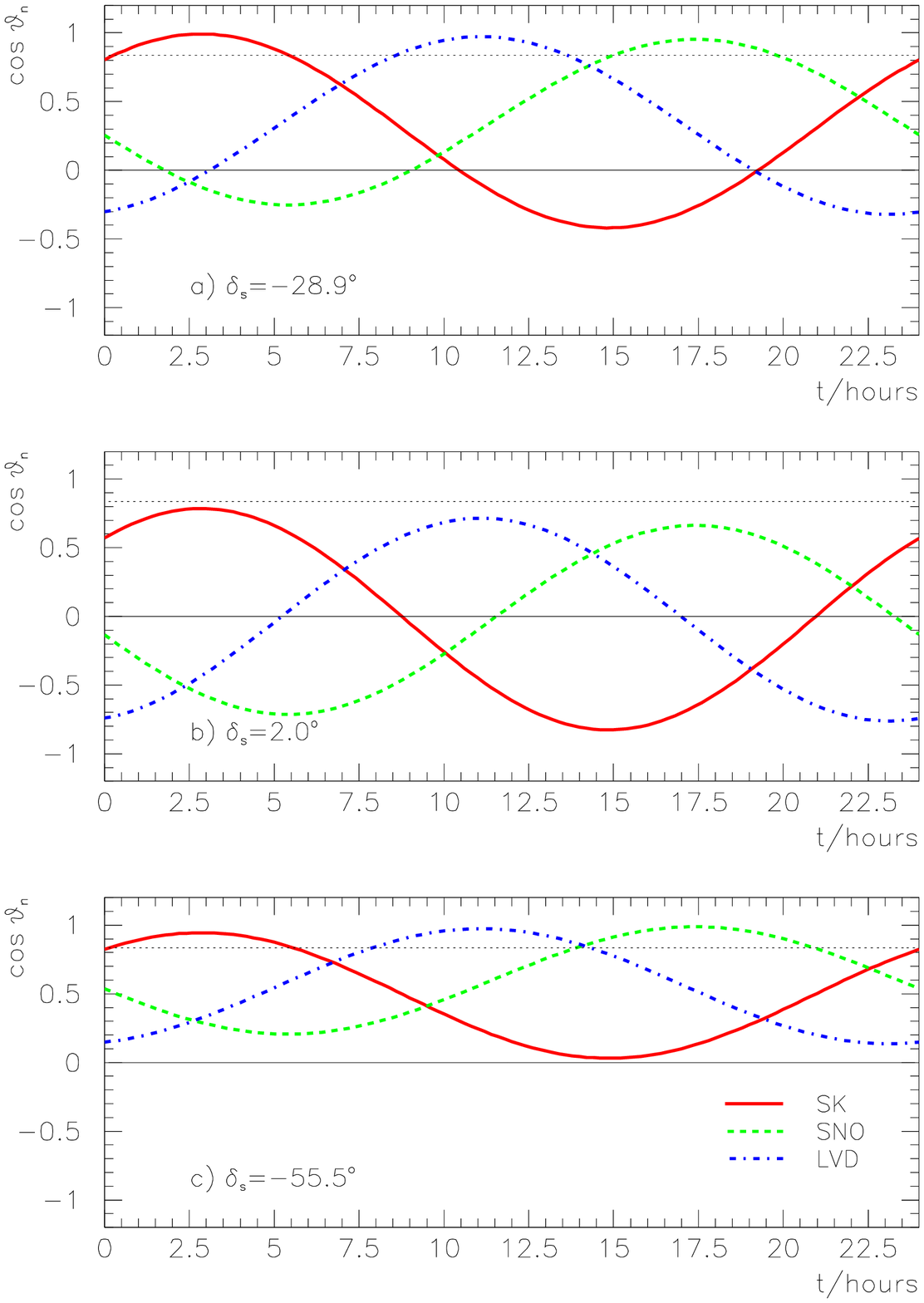, width=14truecm}
\end{center}
\caption{The cosines of the nadir angles $\theta_n$ of SuperKamiokande, SNO and LVD detectors with respect to the supernova as functions of the arrival time of neutrino burst. The three panels refer to three different locations of the star in the galactic plane (given by the declination angle $\delta_s$). We fixed $t=0$ as the time at which the star is aligned with the  Greenwich meridian.
} 
\label{fig:nadir} 
\end{figure}

\begin{figure}[hbt]
\begin{center}
\epsfig{file=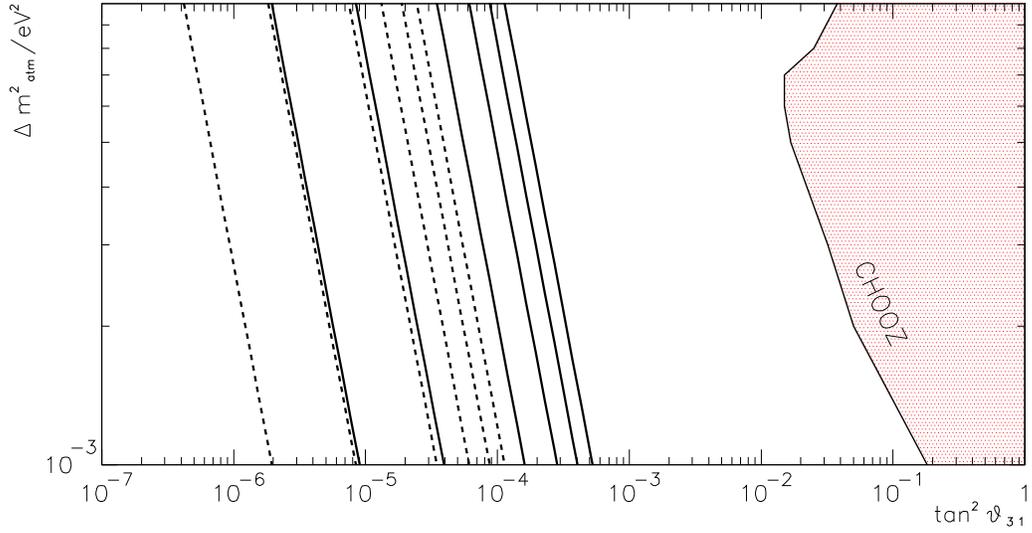, width=15truecm}
\end{center}
\caption{Lines of constant flip probability $P_H$ in the $\Delta m^2_{atm}$-$\tan^2 \theta_{3 1}$ plane. The solid lines refer to $E=50$ MeV and correspond, from right to left, to $P_H=0.05,0.1,0.2,0.4,0.8,0.95$. The dashed lines correspond to the same values of $P_H$ with $E=5$ MeV.  The exclusion region from the CHOOZ experiment is shown.   
} 
\label{fig:ph} 
\end{figure}

\begin{figure}[hbt]
\begin{center}
\epsfig{file=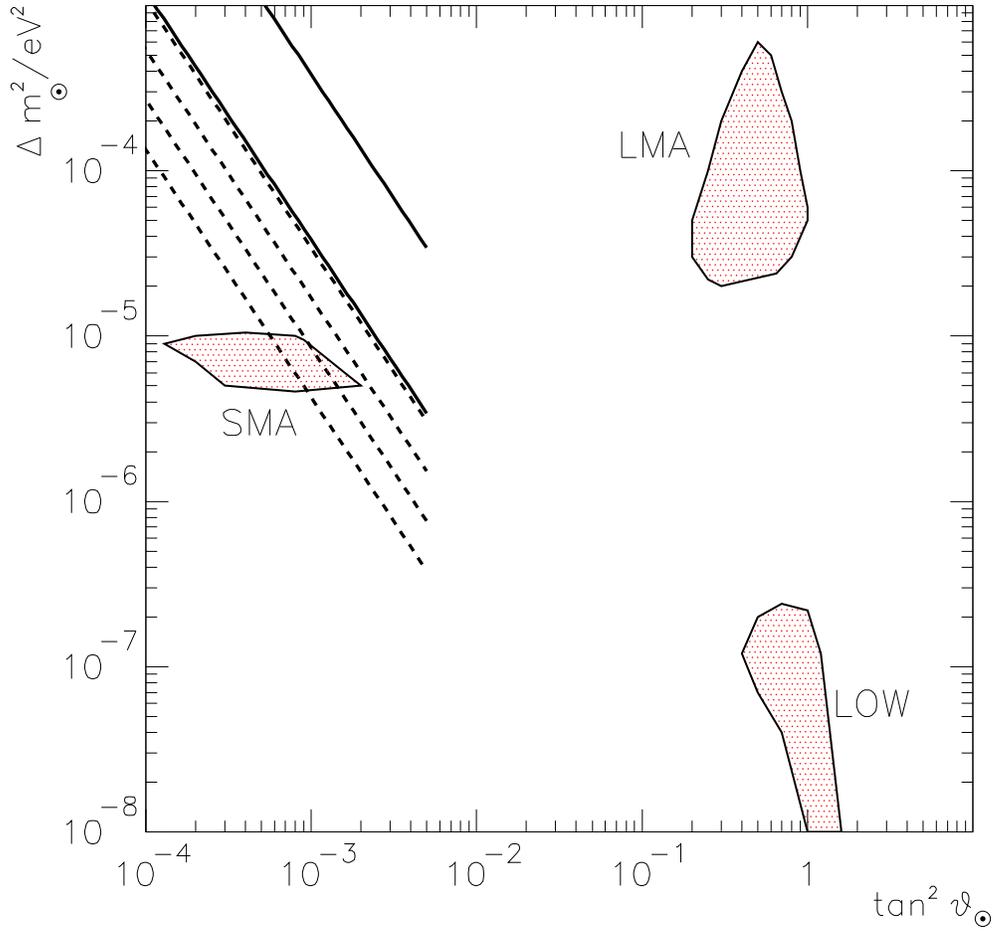, width=15truecm}
\end{center}
\caption{Lines of constant jump probability in the low density resonance, $P_L$, in the $\tan^2 \theta_{\odot}$-$\Delta m^2_{\odot}$ plane. The solid lines correspond to $P_L=0.05$, $E=50$ MeV (upper line) and $E=5$ MeV (lower line). The dashed lines correspond to $P_L=0.5$ and, from the upper to the lower, $E=40,20,10,5$ MeV. The plot of the lines is restricted to the region of parameters for which the L-resonance inside the star occurs at densities larger than $1~{\rm g\cdot cm^{-3}}$ (see sect. \ref{sec:2.1} of the text). The allowed regions for the SMA, LMA and LOW solutions of the solar neutrino problem are represented.} 
\label{fig:adiablow} 
\end{figure}

\begin{figure}[hbt]
\begin{center}
\epsfig{file=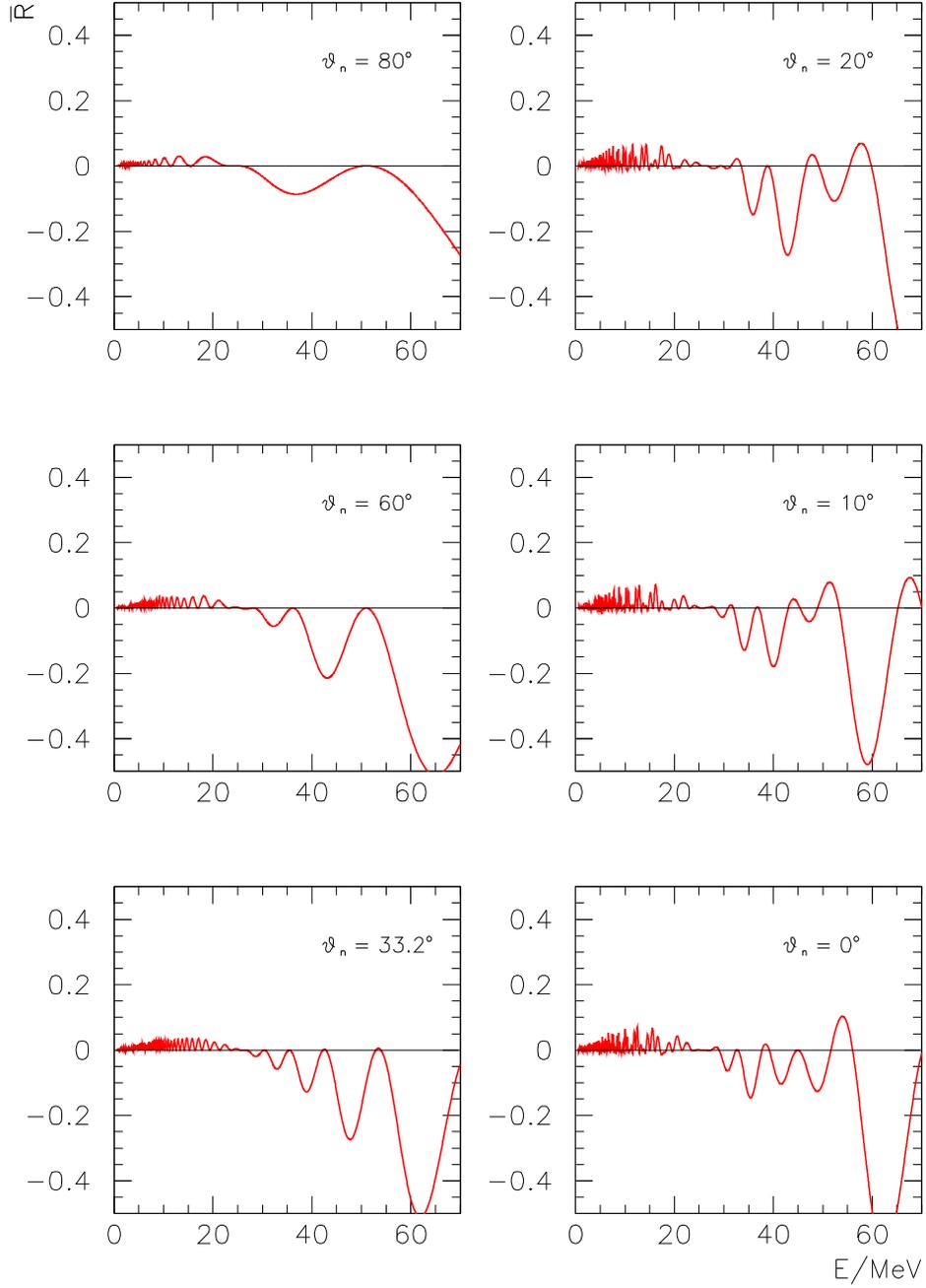, width=14truecm}
\end{center}
\caption{The relative Earth matter effect in $\barnue$ channel, ${\bar R}$, as function of the antineutrino energy for LMA oscillation parameters and various values of the nadir angle $\theta_n$. We have taken $\Delta m^2_{\odot}=5\cdot 10^{-5}~{\rm eV^2}$, $\sin^2 2\theta_{\odot}=0.75$;  $T_{\bar e}=5$ MeV, $T_x=8$ MeV. The figure refers to normal mass hierarchy (or inverted hierarchy with $P_H=1$).   
} 
\label{fig:fig5} 
\end{figure}

\begin{figure}[hbt]
\begin{center}
\epsfig{file=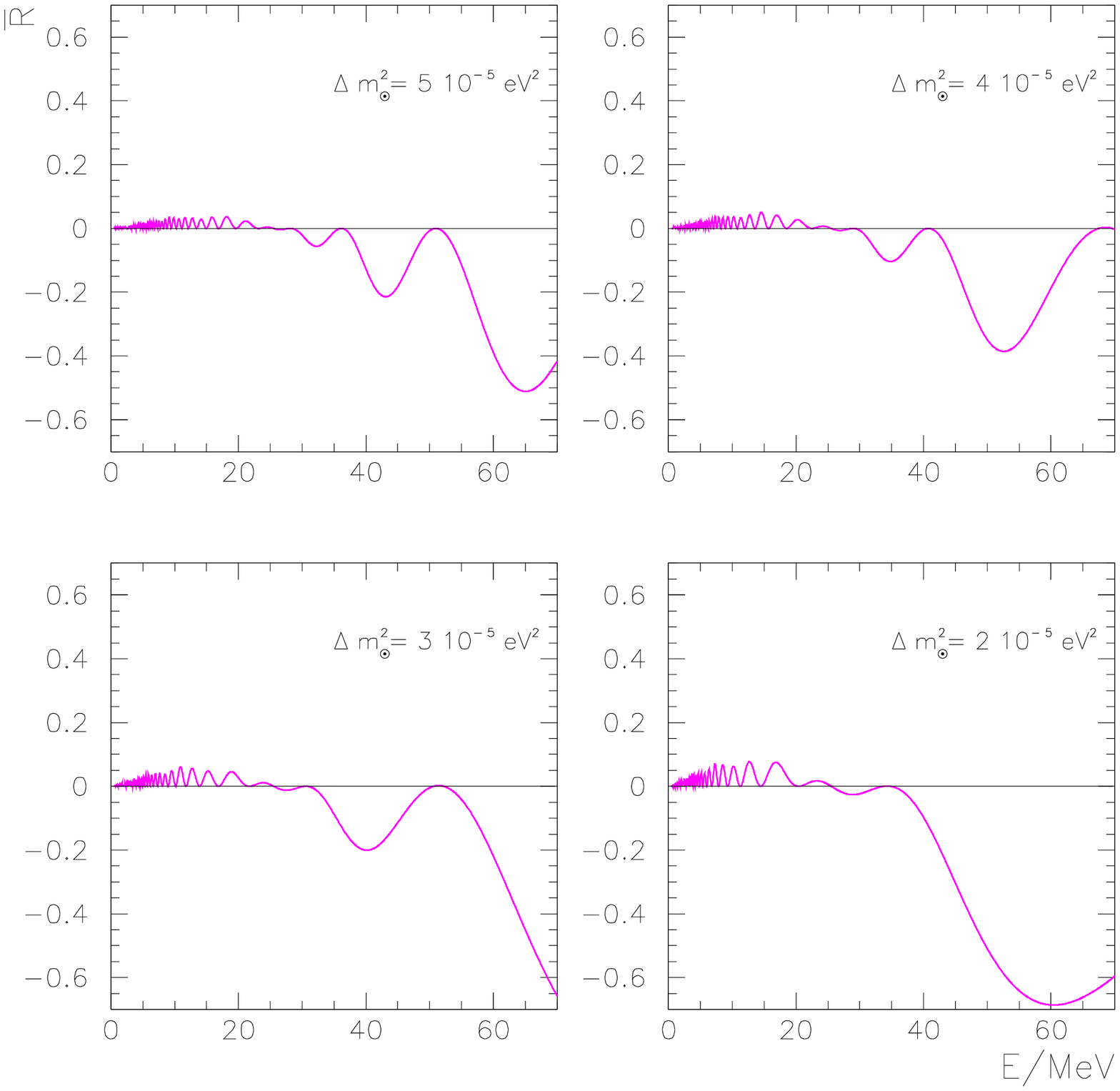, width=15truecm}
\end{center}
\caption{The same as fig. \ref{fig:fig5} for $\theta_n=60^\circ$ and various values of $\Delta m^2_{\odot}$.   
} 
\label{fig:fig7} 
\end{figure}

\begin{figure}[hbt]
\begin{center}
\epsfig{file=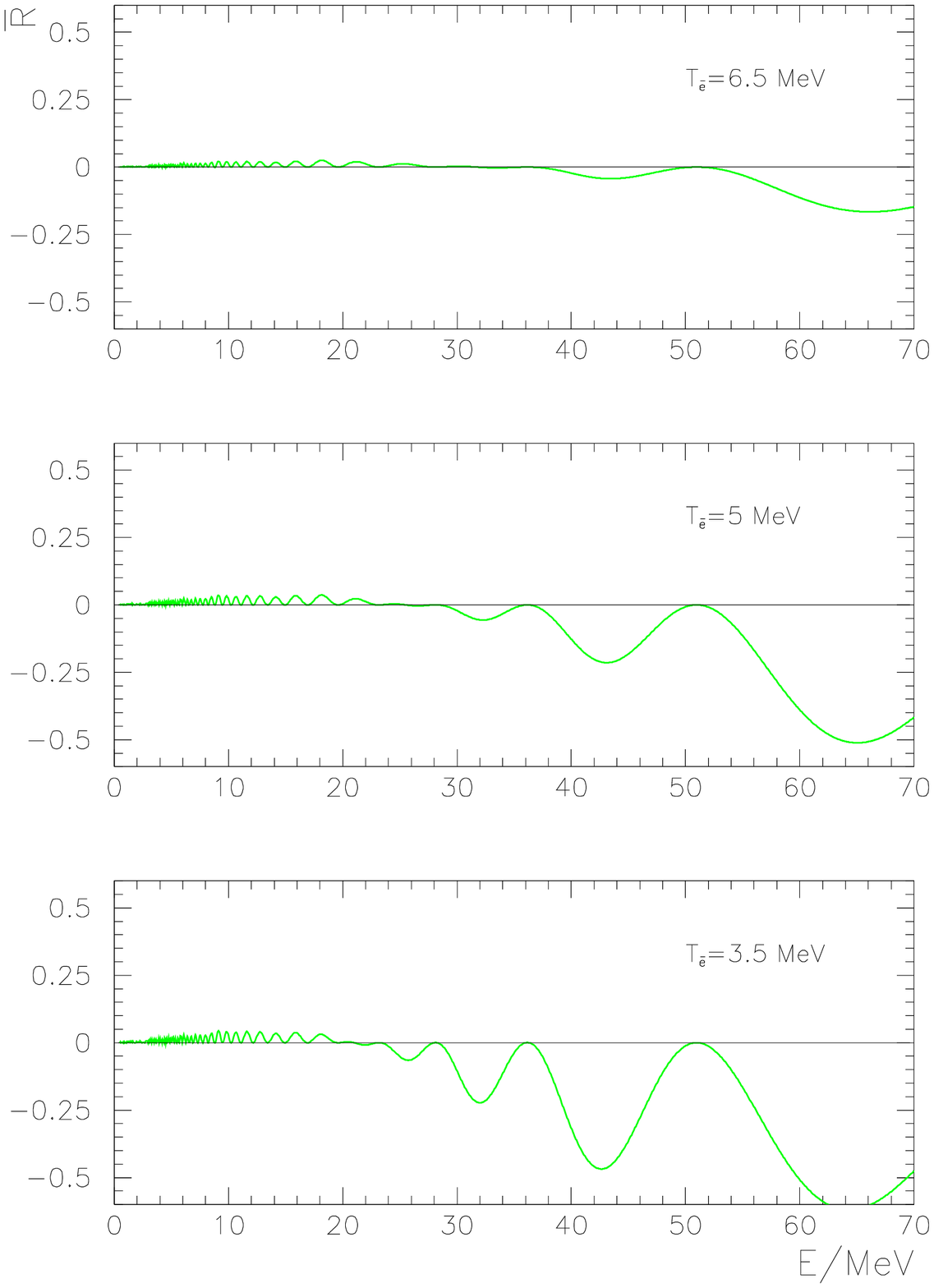, width=15truecm}
\end{center}
\caption{The same as fig. \ref{fig:fig5} for $\theta_n=60^\circ$ and various values of $T_{\bar e}$.   
} 
\label{fig:fig8} 
\end{figure}

\begin{figure}[hbt]
\begin{center}
\epsfig{file=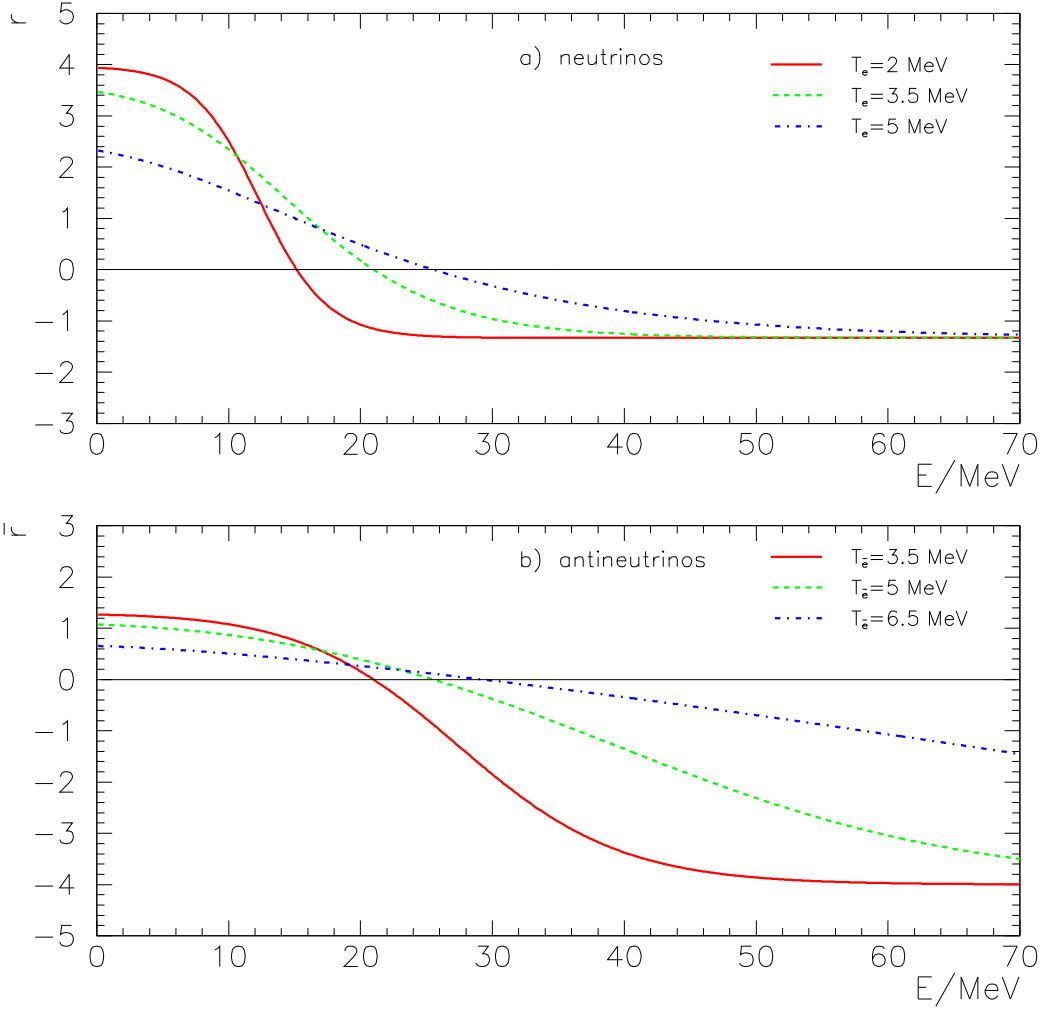, width=15truecm}
\end{center}
\caption{The flux factors $r$ and $\bar r$ for various values of the temperatures $T_e$ and $T_{\bar e}$. We have taken $T_x=8$ MeV, $\Delta m^2_{\odot}=5\cdot 10^{-5}~{\rm eV^2}$, $\sin^2 2\theta_{\odot}=0.75$ and $P_H=1$.     
} 
\label{fig:fluxf} 
\end{figure}

\begin{figure}[hbt]
\begin{center}
\epsfig{file=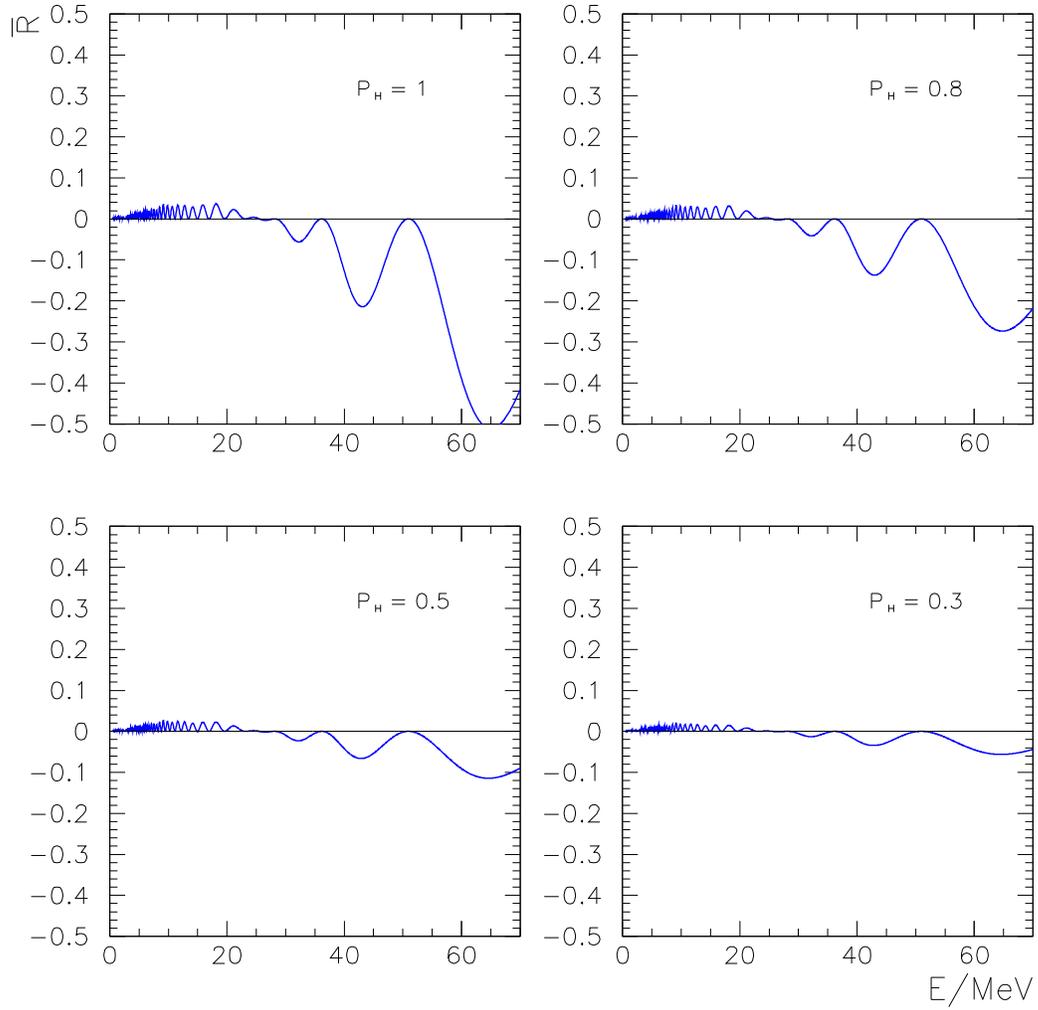, width=15truecm}
\end{center}
\caption{The same as fig. \ref{fig:fig5} for inverted mass hierarchy and 
various values of $P_H$. We have taken $\theta_n=60^\circ$.   
} 
\label{fig:fig6} 
\end{figure}

\begin{figure}[hbt]
\begin{center}
\epsfig{file=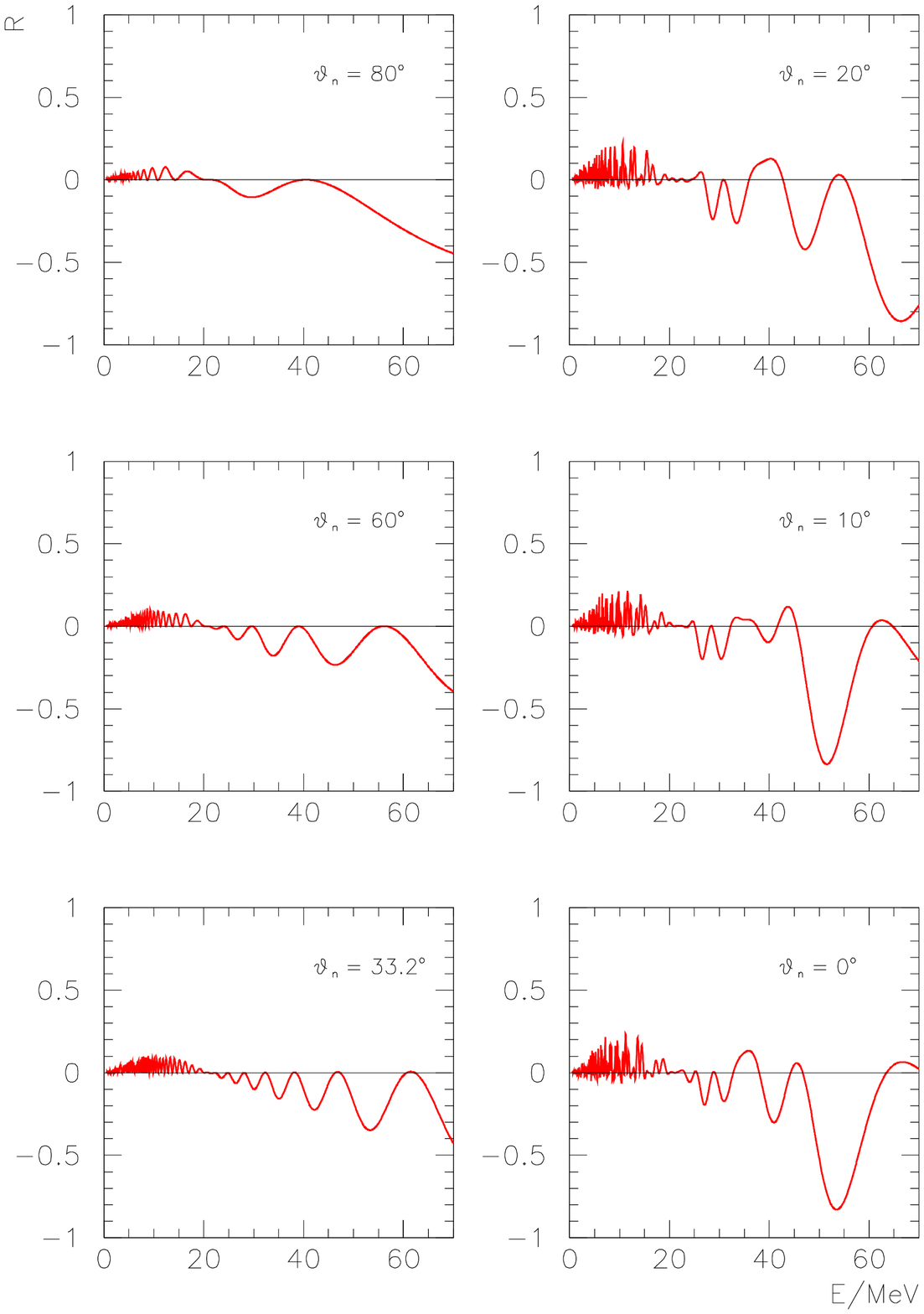, width=14truecm}
\end{center}
\caption{The relative Earth matter effect in $\nu_e$ channel, $R$, as function of the neutrino energy for LMA oscillation parameters and various values of the nadir angle $\theta_n$. We have taken $\Delta m^2_{\odot}=5\cdot 10^{-5}~{\rm eV^2}$, $\sin^2 2\theta_{\odot}=0.75$;  $T_e=3.5$ MeV, $T_x=8$ MeV; $P_H=1$ (or inverted hierarchy).   
} 
\label{fig:fig1} 
\end{figure}

\begin{figure}[hbt]
\begin{center}
\epsfig{file=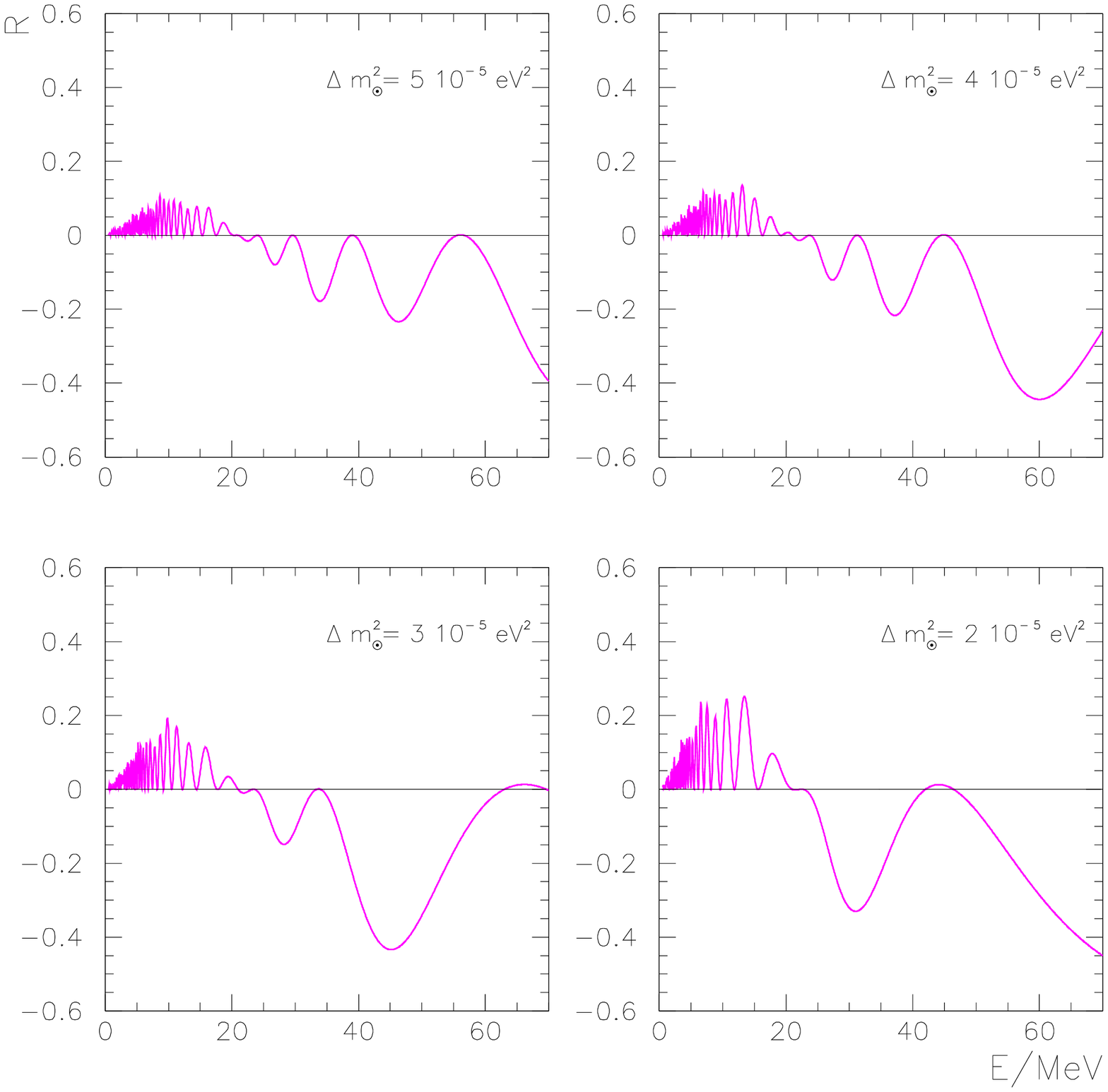, width=15truecm}
\end{center}
\caption{The same as fig. \ref{fig:fig1} for $\theta_n=60^\circ$ and various values of $\Delta m^2_{\odot}$.   
} 
\label{fig:fig3} 
\end{figure}

\begin{figure}[hbt]
\begin{center}
\epsfig{file=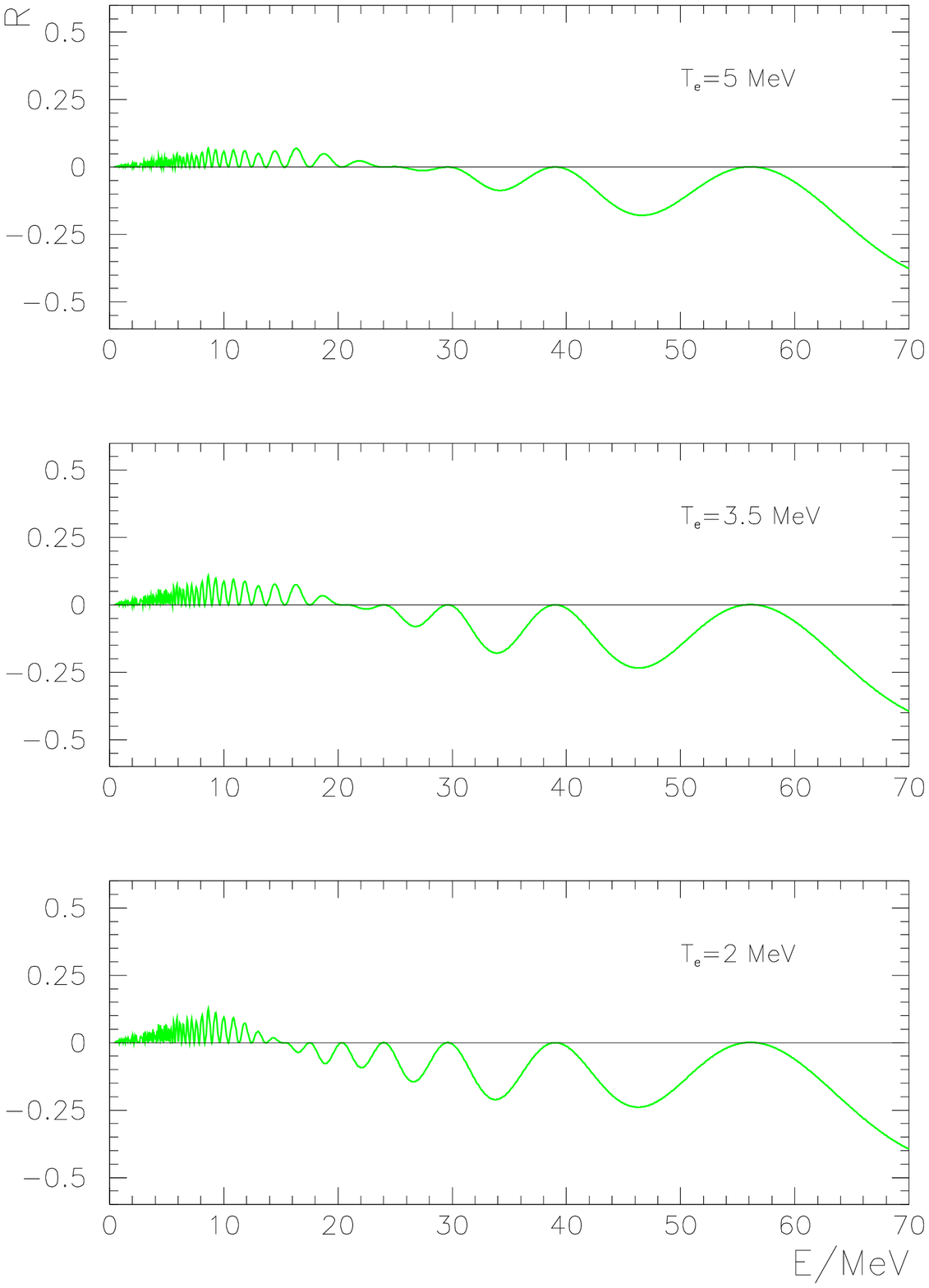, width=15truecm}
\end{center}
\caption{The same as fig. \ref{fig:fig1} for $\theta_n=60^\circ$ and various values of $T_e$.   
} 
\label{fig:fig4} 
\end{figure}

\begin{figure}[hbt]
\begin{center}
\epsfig{file=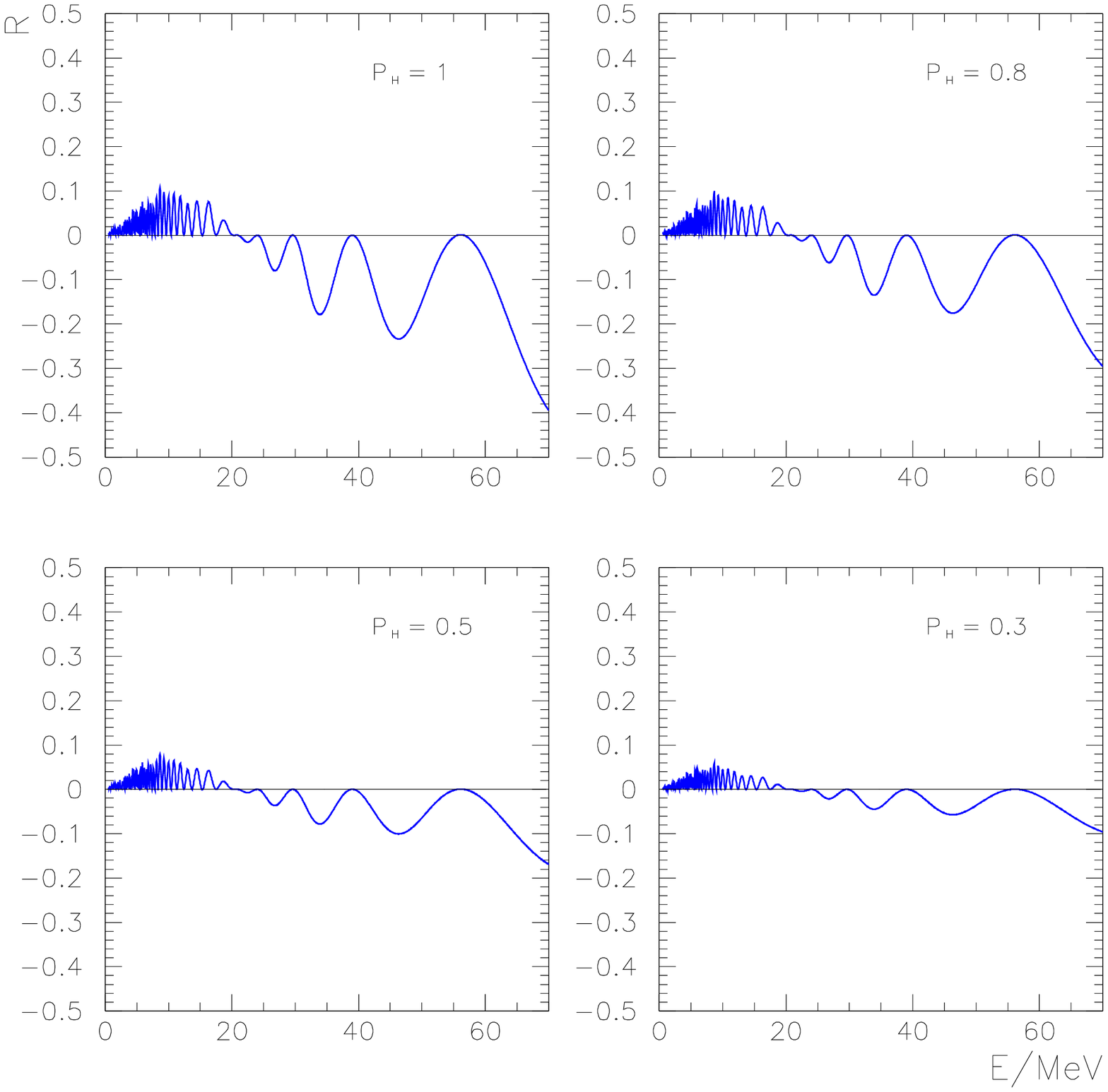, width=15truecm}
\end{center}
\caption{The same as fig. \ref{fig:fig1} for $\theta_n=60^\circ$ and various values of $P_H$.   
} 
\label{fig:fig2} 
\end{figure}

\begin{figure}[hbt]
\begin{center}
\epsfig{file=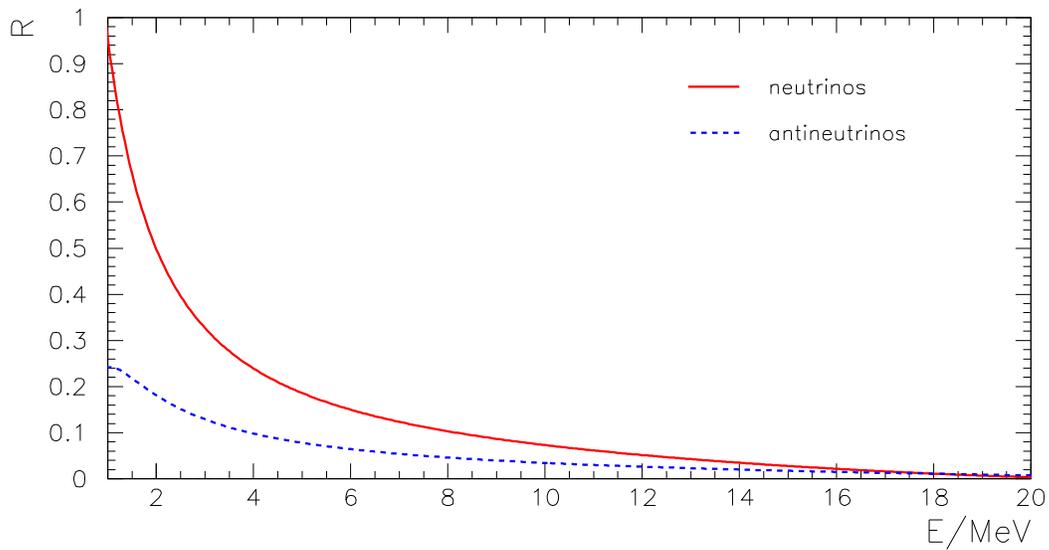, width=15truecm}
\end{center}
\caption{The relative deviations $R$ (solid line) and $\bar R$ (dashed line) as functions of the neutrino (antineutrino) energy for LOW oscillation parameters. We have taken $\Delta m^2_{\odot}=10^{-7}~{\rm eV^2}$, $\sin^2 2\theta_{\odot}=0.9$;  $T_e=3.5$ MeV, $T_{\bar e}=5$ MeV, $T_x=8$ MeV and
 $\theta_n=25^\circ$. We have also assumed $P_H=1$ and $P_L=0$.   
} 
\label{fig:fig10} 
\end{figure}

\begin{figure}[hbt]
\begin{center}
\epsfig{file=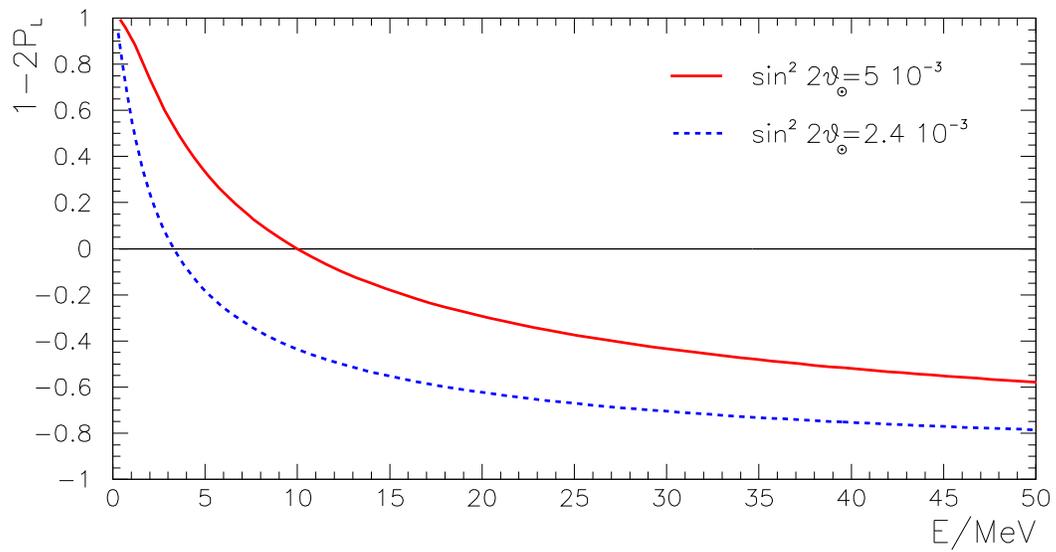, width=15truecm}
\end{center}
\caption{The quantity $1-2P_L$ as a function of the neutrino energy for $\Delta m^2_{\odot}=6\cdot 10^{-6}~{\rm eV^2}$ and different values of $\sin^2 2\theta_{\odot}$ in the SMA region. We have taken $C=4$  in the profile (\ref{eq4}).  
} 
\label{fig:pl} 
\end{figure}

In the figs. \ref{fig:histt1}-\ref{fig:histt17} we show examples of the spectra expected at SK, SNO and LVD for oscillation parameters from the LMA solution, $P_H=1$ and various arrival times of the neutrino burst.  
We considered a supernova located in the direction of the galactic center (fig. \ref{fig:nadir} a)) at a distance $D=10$ Kpc and releasing a total energy $E_B=3\cdot 10^{53}$ ergs.
The histograms represent the numbers of events from the reaction (\ref{antinu_r}) for SK (panels a)) and LVD (panels b)); 
the panels c) show the sum of the numbers of events from the reactions (\ref{antinu_rd}) and  (\ref{antinu_r}) at SNO. In  d) we plot the numbers of events in the inner volume of SNO from the scattering (\ref{nu_rd}).  

As can be seen in fig. \ref{fig:nadir} a), for $t=1$ hour the neutrinos arriving at SK have core crossing trajectory ($\cos \theta_n=0.93$). For SNO the trajectory crosses the mantle only and is rather superficial ($\cos \theta_n=0.10$); the LVD detector is not shielded by the Earth. 
The corresponding spectra are shown in fig. \ref{fig:histt1} for $\Delta m^2_{\odot} =5\cdot 10^{-5}~{\rm eV^2}$ and $\sin^2 2\theta_{\odot}  =0.75$ and the same temperatures as in the figs. \ref{fig:fig5} and \ref{fig:fig1} (see section \ref{sec:2.1}).
The spectrum of SK events exhibits deviations from the undistorted spectrum in some isolated bins, which correspond to the minima in  the antineutrino spectrum.
At SNO the Earth effect produces a narrowing of the spectrum. The difference with respect to the SK spectrum is attributed to the smaller oscillation phase (shorter trajectory in the Earth) of the neutrinos arriving at SNO.

The comparison of the results of the three experiments can be performed in various ways. If the direction to the supernova is determined, the LVD spectrum is known to be free from regeneration effects. Therefore it can be used to predict the energy distributions at SK and SNO without Earth matter effects.  Such predictions can be compared to the observations of SNO and SK. 
Due to the relatively small statistics  of the LVD events, however, the accuracy of the
reconstruction will not be high  and  the deviation from the
undistorted spectrum (e.g. in the range 40 - 65 MeV) will not be 
larger than 2$\sigma$.

Higher statistical significance is obtained if data from a second large volume detector are available. Kilometer-scale neutrino telescopes, though primarily devoted to the study of high-energy neutrinos, are expected to be sensitive to supernova neutrinos. In particular, for a supernova at distance $D=10$ Kpc the  ice Cerenkov detector of AMANDA \cite{Spiering:2001qd} would observe  more than $2\cdot 10^{4}$ events from $\barnue$ scattering on protons, eq. (\ref{antinu_r}) \cite{Halzen:1996ex,amanda}. Unfortunately, the presence of a relatively large background and the absence of sensitivity to the neutrino energy spectrum \cite{Halzen:1996ex,amanda,privcomAM} strongly restrict the potential of the study of this signal. We mark, however, that substantial upgrades of the experimental apparatus are possible \cite{privcomAM} and the optimization of the detector for supernova neutrino observation would be of great interest.
Besides the present neutrino telescopes, the detection of supernova neutrinos is among the goals of future large volume detectors, like UNO \cite{Jung:1999jq} and NUSL \cite{nusl}.  We find that the comparison of the energy spectra observed by SK and by another detector with comparable or larger statistics could establish the Earth matter effects at more than $\sim 5 ~\sigma$ level.

Even larger regeneration effect can be realized due to specific features of the original neutrino spectra and of the solution of the solar neutrino problem.  The figure \ref{fig:histopt} shows the same spectra as fig. \ref{fig:histt1} for $\Delta m^2_{\odot} =3\cdot 10^{-5}~{\rm eV^2}$, $\sin^2 2\theta_{\odot}  =0.9$, $T_e=3$ MeV and $T_{\bar e}=4$ MeV.  In this case the regeneration effect is substantially larger (see also figs. \ref{fig:fig7} and \ref{fig:fig8}) and can be established by SK-LVD comparison with high ($\gta 3 ~\sigma$) statistical significance.
Notice also that the statistics of each experiment, and therefore the power of the comparison between different detectors, is higher for smaller distance $D$ to the supernova and/or  larger binding energy $E_B$.
For instance for $D=3$ Kpc and $E_B\simeq 4.5 \cdot 10^{53}$ ergs the statistics is $\sim 17$ times higher and the differences between the spectra of SK and LVD (unshielded) can be as large as $(6-10)~\sigma$.
\\

\noindent
Besides the comparison of the spectra, more specific criteria of identification of the Earth effect can be elaborated if the location of the supernova and the solar neutrino oscillations parameters are known. For instance, for LMA parameters and rather superficial trajectory in the mantle the effect consists in a narrowing of the spectrum (see e.g. figs. \ref{fig:histt1}-\ref{fig:histopt} panels c)-d)). Thus the comparison of the widths of the spectra at different detectors may establish the Earth effect. 

As a further illustration, in figs. \ref{fig:histt8}  and \ref{fig:histt17} we show the expected spectra for the same parameters as in fig. \ref{fig:histt1} but different arrival times of the signal (see fig. \ref{fig:nadir} a)).
For $t=8$ hours SNO is unshielded while SK and SNO have deep trajectories in the mantle. For $t=17$ hours SNO has core crossing trajectory  ($\cos \theta_n\simeq 1$) and LVD observes effects of regeneration in the mantle only; SK is unshielded.

For neutrino parameters from the SMA solution the Earth effects on the spectra are smaller than $\sim 5-10\%$. 
This is explained by the fact that the narrow resonance peaks appear in the low energy part of the spectrum where  the smoothing effect of the integrations (\ref{eq:dnem}) is stronger, thus suppressing the effect. Moreover, in the low energy region the detection cross sections and efficiencies are small.
Therefore, regeneration effects with SMA parameters appear difficult to be observed.  Similar conclusions hold for the case of LOW parameters, for which the effect is small and localized at low energies.  
\\

As we have mentioned in sect. \ref{subsec:invhier},  already 
the very  fact of establishing the
Earth matter effect in the neutrino and/or in antineutrino channel will
have important implications for the neutrino mass and flavor spectrum. 
For this it will be  enough to study some  integral effect of
regeneration. 
Let us consider the following possibility. 
As we have discussed in sect.~\ref{sec:4}, for LMA parameters the regeneration effect is negative above  the
critical energy  and moreover the relative size of the 
effect increases with $E$. The Earth matter effect at low energy is small
mainly due to the dependence of the regeneration factor 
$f_{reg} \propto E$ (eq. (\ref{depth})). Therefore to identify the regeneration effect 
one can compare the signals from the process (\ref{antinu_r}) in the various detectors at low, $E < E_s$, and high,  
$E >  E_s$, energies \cite{Takahashi:2001ep}, where $E_s$ is some separation energy. 
Let us introduce the numbers of events
\be
N_L \equiv 
\int_{E_{th}}^{E_s} dE_e
{dN({\bar \nu}p) \over dE_e}~, ~~~~~~ 
N_H \equiv \int_{E_s}^{\infty} dE_e
{dN({\bar \nu}p) \over dE_e}~
\ee
and the ratio
\be
{\cal R} \equiv \frac{N_L}{N_H}~.
\label{ratio}
\ee
In absence of Earth effects ${\cal R}$ has the same value for every experiment provided that the detection efficiencies are independent of energy at $E>E_{th}$, or in the particular case of equal efficiencies. Therefore, differences in the quantity (\ref{ratio}) are entirely due to regeneration effects: stronger effect corresponds to larger  ${\cal R}$.
In fig.~\ref{fig:ratios}  we show  the dependence of ${\cal R}$ 
on the separation energy $E_s$ for different detectors; we have taken $E_{th}=20$ MeV.
The panels a) and b) of the figures refer to the situations illustrated in  figs. \ref{fig:histt1}  and \ref{fig:histopt} respectively.
In agreement with the analyses of the spectra, in the latter case (panel b)) the effect is stronger: the deviation of the ratio ${\cal R}$ for SK from the value ${\cal R}_0$ in absence of regeneration, given by LVD, can be as large as $(2- 3) ~\sigma$. Larger deviation ($(4- 5) ~\sigma$) is realized if ${\cal R}_0$ is provided by an experiment with volume comparable or larger than the one of SK.

\section{Discussion}
\label{disc}

Within 2 - 3 years the solution of the solar neutrino problem can be identified by the
results of the SNO, KamLAND and  BOREXINO experiments. In particular,
KamLAND will be able to establish the LMA solution and to measure
$\Delta m^2_{\odot}$ 
and $\sin^2 2\theta_{\odot}$  with 10 - 20\% accuracy \cite{Barger:2001hy,Barbieri:2000sv,Murayama:2000iq,Piepke:2001tg}. 
This will enormously sharpen the predictions for the Earth matter effects. 

The possibility exists that $U_{e3}$ will be determined 
by MINOS \cite{Wojcicki:2001ts} provided that its value is not too far from 
the present upper bound, eq. (\ref{ue3}). In this case, $U_{e3}$ is certainly in the 
adiabatic range.   
In a long perspective, a neutrino factory \cite{Albright:2000xi} 
will be able to cover the whole the range of $U_{e3}$ relevant for 
supernova neutrinos. Thus, either $U_{e3}$ will be measured 
or the upper bound on $U_{e3}$ will be so strong that 
the transition in the supernova will be certainly non-adiabatic.\\

Let us consider possible implications of the supernova 
neutrino results depending on the solution 
of the solar neutrino problem.\\

\noindent
1). Suppose that the LMA solution will be identified with parameters 
close to the present best fit point 
($\Delta m^2_{\odot} = (3 - 6)\cdot  10^{-5}$ eV$^2$). As we have seen, 
in this case the regeneration effect can be 
observed.  

The features of the Earth matter effects depend on the value of $U_{e3}$ and on the type of mass hierarchy. 
For normal mass hierarchy and $U_{e3}$ in the adiabatic range 
(which appears as the most plausible scenario) we expect 
 regeneration effects in the antineutrino channel and 
no  effect in the neutrino channel (see sect. \ref{subsec:invhier}). 
In the supernova data further confirmations of
such a possibility  are (i) the absence of the
neutronization peak in $\nu_e$ and appearance of the 
$\nu_{\mu}/\nu_{\tau}$ neutronization peak, (ii) hard spectrum of 
$\nu_e$  during the cooling stage: $\langle  E_e  \rangle > \langle  E_{\bar{e}}  \rangle$.
 
The relative size of the effect in $\barnue$ channel, $\bar R$, is determined by the regeneration factor and the flux factor, according to eq. (\ref{rat2barnu}). 
At high energies, when the flux factor reaches the asymptotic  
value (\ref{banti_h}), eq. (\ref{rat2barnu}) gives: 
\be
{\bar R}(E \gg {\bar E}_c) \sim -\frac{{\bar f}_{reg}}{\sin^2 \theta_{\odot} }~.  
\label{rel-lma-as}
\ee 
That is, the effect is completely predicted in terms of solar 
oscillation parameters. 
\\

\noindent 
In the case of inverted mass hierarchy the Earth matter effect 
should be observed in the neutrino channel and no effect is 
expected in the antineutrino channel if $|U_{e3}|^2 >  10^{-5}$ (see fig. \ref{fig:ph}). 
This possibility will be confirmed by the observation of the 
$\nu_e$-neutronization peak and of an hard spectrum of the 
$\bar{\nu}_e$ during the cooling stage. 
The relative size of the Earth matter effect, $R$, is given by eq. (\ref{redrat}) (with the replacement $P_H\rightarrow 1$) with the 
 high energy  asymptotic
(see eq. (\ref{Rinf}))
\be
R(E \gg E_c)\sim -\frac{f_{reg}}{\cos^2 \theta_{\odot} }~.
\label{rel-lma-4}
\ee       
\\

\noindent
In the limit $|U_{e3}|^2 \ll  10^{-5}$ the high density resonance is
inoperative, so that the result is insensitive to the mass hierarchy. 
Oscillations appear in both the neutrino and 
antineutrino channels, and, at high energies, they are determined by the solar oscillation
parameters (eqs. (\ref{rel-lma-as}) and (\ref{rel-lma-4})). 
The ratio of the relative effects at high energies equals:
\be
\frac{R}{{\bar R}} 
\simeq \tan^2 \theta_{\odot}  \frac{f_{reg}}{\bar{f}_{reg}}~.        
\label{rel-ratio}
\ee 
So, possible checks of this  equality  would be  the confirmation 
of the neutrino scheme with very small $U_{e3}$. \\

\noindent
If  $U_{e3}$ is in the intermediate 
region: $|U_{e3}|^2 \sim  10^{-6}- 10^{-5}$ the situation is more complicated. 
One expects to observe oscillations both in the 
neutrino and antineutrino channels; the regeneration effect depends on the mass hierarchy
and on the specific value of  $U_{e3}$. 
In the case of normal mass hierarchy, the relative effect in the
neutrino channel
is proportional to $P_H = P_H (U_{e3})$ (eq. (\ref{redrat}))  
and at high energies, when the flux factor reaches the asymptotic 
value we get, (eq. (\ref{Rinf})):
\be
R(E \gg E_c) \sim   - \frac{P_H}{1 - P_H \sin^2 \theta_{\odot} } f_{reg} ~.
\label{rel-lma-na-a}
\ee  
Since $f_{reg}$ is determined by the solar parameters, by
measuring the relative deviation $R$ we can determine the  value of
$P_H$ via the eq. (\ref{rel-lma-na-a}) and therefore get information about $U_{e3}$. 

For antineutrinos with inverted hierarchy we find the  high energy asymptotic (see eq. (\ref{Rbarnuinf})): 
\be
{\bar R}(E \gg {\bar E}_c)\sim -  \frac{P_H} {1 -  P_H \cos^2 \theta_{\odot} } {\bar f_{reg}} ~.
\label{rel-lma-anta}
\ee  
\\

\noindent
In practice,  the observation of the Earth matter effect in the 
$\bar{\nu}_e$ channel and absence of the effect in $\nu_e$
channel will testify  for normal mass hierarchy and 
$|U_{e3}|^2 > 10^{-5}$.  In the opposite situation,  
 effect in the $\nu_e$ channel and absence of the effect in $\bar{\nu}_e$
channel, the  inverted  hierarchy will be identified with 
$|U_{e3}|^2 > 10^{-5}$.  
However the  present experiments have lower sensitivity to $\nu_e$ fluxes with respect to the fluxes of $\barnue$,
 so that it may  be 
difficult to establish ``zero"  regeneration effect with high 
enough accuracy. 

If  the Earth matter effect is observed in both channels,  
one should   compare the size of the effect with that predicted in the 
absence of the high resonance in a given channel. 
Thus, if the observed signal in the neutrino channel is smaller than what
is predicted in the assumption of $P_H = 1$, 
whereas in the antineutrino channel prediction 
and observation coincide, we will conclude that 
the hierarchy is normal and the ratio of the 
observed to predicted signals in the neutrino channel can give the
value of $P_H$. The opposite case of coincidence of 
the predicted and observed signals in the neutrino channel 
and suppressed  observed signal in the antineutrino channel 
will testify for the inverted mass hierarchy. 

Besides the probing of the neutrino mass spectrum and mixing, a study of the properties of the original neutrino fluxes can be done with Earth matter effects. In principle, a detailed study of the observed energy spectra will allow to reconstruct the flux factor as well as to determine  the critical energy $E_c$.\\

\noindent
2). Suppose that the future 
solar neutrino experiments will identify the SMA solution. 
In this case, the Earth matter effect is expected in the 
neutrino channel only, and only if the high density resonance
is inoperative. This requires the inverted mass hierarchy or 
very small $U_{e3}$ in the case of normal mass hierarchy. 
As discussed in sect. \ref{subsec:sma}, the effect is small
and   difficult to be observed  
even in the most favorable situations. 

For the rather  plausible case of the normal mass hierarchy and 
$|U_{e3}|^2 >   10^{-5}$  no Earth matter effect should be seen. 

The observation of the Earth matter effect in the SMA case will allow 
to conclude that the mass hierarchy is inverted or 
the hierarchy is normal but $U_{e3}$ is very small: 
$|U_{e3}|^2 \ll  10^{-5}$. 
In principle, the intermediate  case $|U_{e3}|^2 \sim  10^{-5}$ can be identified 
if the observed signal will be smaller that the expected one for 
$P_H = 1$. Notice that from eqs. (\ref{rat2}) and (\ref{red}) one gets
the low energy asymptotic for the flux factor:  
\be
R(E \ll E_c)=
 P_H (1-2P_L) f_{reg} \frac{1}{P_H P_L} =  f_{reg}(1-2P_L) \frac{1}{ P_L}~, 
\label{rel-lma-ant5}
\ee   
which does not depend on $P_H$. So it will be difficult to  
disentangle the effect of  $P_H$ from 
uncertainties in the original neutrino fluxes. 
\\
 
The situation can be much more complicated if a sterile neutrino 
exists. This can be clarified in 2 - 3 years: 
the MiniBooNE experiment and further searches  for sterile 
neutrinos in the solar and atmospheric neutrino experiments will allow 
to establish the existence of sterile neutrinos. 

Negative results of the searches will strongly favor the $3\nu$ schemes 
discussed in this paper. Still some uncertainty will remain: 
sterile neutrinos, unrelated to the LSND result, may exist and weakly 
mix with active
neutrinos. Even a very small mixing (unobservable by other means)
of  sterile states 
with masses in the wide range from sub eV   up to 10 keV  can strongly
modify the properties of the neutrino  burst. 

\section{Conclusions}
\label{concl} 

\noindent 
1). There is a big chance that at least one of the existing detectors (SK,
SNO, LVD) will be
shielded by the Earth at the moment of  arrival of a supernova neutrino
burst, so that the Earth matter effect on the neutrino flux will be 
observed. For supernova in a region close
to the galactic center the most plausible configuration is that 
for two detectors the trajectories of the neutrino burst cross the Earth, whereas 
the third detector is unshielded. 

We found that the detectors considered can register Earth matter effects 
($\cos\theta_n> 0 $) for a significant fraction ($\gta 60\%$)
of the possible arrival times of the signal  and core effect for $\sim 20\%$ of the times. 
These fractions may be even larger depending on the specific location 
of the star in the galaxy. 

The comparison of the signals  from different detectors  
allows to identify the Earth matter effect and to 
get information on the neutrino  mass spectrum,  substantially reducing the
uncertainties related to the model of the star and to the original neutrino
fluxes. 
\\

\noindent 
2). We studied the effects of the matter of the Earth on supernova 
neutrinos in the framework of three flavours with either 
normal or inverted hierarchy.\\

\noindent 
3). The strongest regeneration effect is expected for oscillation
parameters in the region of the LMA solution 
of the solar neutrino problem, especially for the lowest values of  $\Delta m^2_{\odot}$
in this region: 
$\Delta m^2_{\odot} = (2 - 5)\cdot 10^{-5}$  eV$^2$. 

In the $\bar{\nu}_e-$ channel the effect exists in the scheme 
with normal mass hierarchy (ordering of the states) or 
in the scheme with inverted mass hierarchy for 
$|U_{e3}|^2 < 10^{-5}$, when the conversion in the high density resonance 
is non-adiabatic. The effect consists in an oscillatory modulation of the 
energy spectra and is negative (except in small energy 
intervals for core crossing trajectories)
 above the critical energy 
$E_c \sim 25$ MeV, thus suppressing the signal. The relative size of the
effect increases with energy and at $E \sim 60 - 70$ MeV it can reach 
$50 - 70\%$. The period of modulation increases with energy and above 
$E \sim 40$ MeV no averaging occurs in the energy 
spectrum of events. 
At low energies the effect  is small and the modulations 
are averaged out. Thus, in the LMA case the most sensitive region 
to the Earth matter effect is above $E\sim 40$ MeV . 
 
The oscillatory picture (position of minima and maxima)  
is very sensitive to $\Delta m^2_{\odot}$.  
For trajectories in the  mantle only the modulation of the spectrum is rather regular. For the 
core-crossing trajectories the structures become 
narrower and  irregular. 
 
The Earth matter effect decreases with the increase of $\Delta m^2_{\odot}$ and  
for $\Delta m^2_{\odot} > 10^{-4}$ eV$^2$ it will be difficult to be observed. 
\\  

\noindent 
4). For the $\nu_e-$ channel substantial Earth matter effect is expected in the
case of normal mass hierarchy, provided that $|U_{e3}|^2 < 10^{-5}$, 
or in the case of inverted mass hierarchy. The effect has 
 oscillatory character, similarly to  the $\barnue-$ case, and  
 can be as large as 100 \% at 
$E > 50 - 60$ MeV. 

Thus, as in the case of antineutrinos, one should   
search for an oscillatory modulation of the signal 
in the energy range $E > 40$ MeV. 

The size of the effect depends on the properties of the original neutrino fluxes via the flux factor. This dependence, however, disappears in the high-energy limit, $E\gta 50$ MeV.
\\

\noindent 
5). For oscillation parameters from the SMA solution the effect appears in the $\nu_e-$ channel only, 
and only if the H-resonance is inoperative
(very small $U_{e3}$ or inverted mass hierarchy). 
The relative effect can be as large as $30\%$ for  core crossing 
trajectories due to the parametric enhancement of oscillations. 
The effect is localized in the low energy part of the spectrum: 
$E = 5 - 10$ MeV. It can be further suppressed by the L-resonance 
factor $(1 - 2P_L)$ which in turn  strongly depends on the density profile
of the star in the outer region. The effect will be strongly 
smoothed in the spectrum of observed events by integrations
over the neutrino energy and  the true energy of electron. 
Practically no distortion is seen and the effect is reduced 
to an increase of the number of events in the region of the peak. 
Effectively this will make the spectrum  narrower.  
One can probably identify this effect by comparing the signals from 
two high statistics experiments.\\ 

\noindent 
6). In the case of LOW solution the Earth matter effect is significant at very low
energies with a  smooth $1/E$ behaviour. If the H-resonance is inoperative, 
in the $\nu_e$ channel the effect (excess of flux)  can reach 
$100 \%$ at $E = 1$ MeV and $20 \%$ in the $\bar \nu_e$ channel
(the difference is mainly due to the difference in the flux factors).  
At $E = 5$ MeV, the effect in the flux is below $(10 - 20) \%$ . 
The nadir angle dependence is  determined by the length  of the trajectory: 
the oscillation length almost equals   the refraction length and depends only weakly on the neutrino energy. For the core crossing trajectories 
the effects can be enhanced due to   parametric effects.\\

\noindent 
7). The identification of the Earth matter effect is possible in a single 
detector (in the LMA case) by the observation of the oscillatory modulation of
the energy spectrum in the high energy part: $E > 40$ MeV. 
Another method consists in the comparison of signals from two 
(or several) different detectors. If one of the detectors is unshielded 
by the Earth its result can be used to reconstruct the  spectrum of the neutrinos arriving at Earth
and make predictions of the signal expected in the  other detectors in absence of matter
effect. 
For a supernova at distance $D = 10$ Kpc and with energy release $E_B=3 \cdot 10^{53}$ ergs we estimated that the Earth matter effects can be established at $(2 - 3) \sigma$ level by comparison between SK, SNO and LVD results, and at $(4 - 5) \sigma$ by comparison between the spectra from two large volume detectors (of SK size or larger).

Another method to identify the Earth matter effect is to study the ratio
of numbers of events in the high and in the low parts of the spectrum 
in different detectors.\\  

\noindent
8). Studies of the Earth matter effect will allow to establish 
or most probably to confirm  the solution of the solar neutrino problem,
to get information about  $U_{e3}$ and to identify the type of  hierarchy  
of the neutrino mass spectrum.

\subsection*{Acknowledgements}
\label{ackn}
The authors are grateful to P.~Antonioli and W.~Fulgione for discussions and for  providing 
material on the LVD experiment. Thanks also to H.~Robertson for giving information on the SNO detector and to
J.~Beacom for helpful comments. 
C.L. wishes to thank A.~Friedland, P.~Krastev, C.~Burgess, A.~Bouchta, C.~De~Los~Heros and R.~Hubbard for useful discussions.


\section*{Appendix A. Parameters of detectors}
\label{appa}
We summarize here the characteristics of the SK, SNO and LVD detectors that have been used in the analysis of section \ref{subsec:6.2}.
\\

For each experiment we consider:
\\

\noindent
1. The position of the detector on the Earth, which is relevant for determining the trajectory of the observed neutrinos inside the Earth (see sect. \ref{sec:2.2}). The locations of the three experiments are given in Table \ref{tab:tab1} in terms of northern latitude, $\delta$,  and eastern longitude  $\alpha$.
\\

\noindent
2. The material which the detector is made of and its fiducial mass. These quantities are quoted in Table \ref{tab:tab1}.
\\

\noindent
3. The detection efficiency ${\cal E}(E'_e)$ (see eq. (\ref{eq:dnem})). 

The SNO efficiency is high, so that the shape of the energy spectrum and the total number of events are determined by the detection cross section \cite{SNOprivcom}. Therefore, we have taken ${\cal E}=1$ in eq. (\ref{eq:dnem}). 

The efficiency of the LVD detector has been provided by the dedicated collaboration \cite{LVDprivcom}. For $E_{th} \leq 10 $ MeV, it is given by the gaussian integral function: 
\begin{equation}
{\cal E} (E_m,E_{th})=\frac{1}{\sqrt{2\pi}} \int_{-\infty}^{x}
e^{-\frac{y^{2}}{2}} dy~,
\label{a1}
\end{equation}
where $x \equiv {(E_m-E_{th})}/{\sigma_{{th}}}$ and $E_m$ is the measured energy of the lepton. The values of the parameters $E_{th}$ and $\sigma_{{th}}$ are: $E_{th} = 4$ MeV and $\sigma_{th} =0.74$  MeV for the core counters (mass $M=0.57$ ktons);  $E_{th} = 7$ MeV and $\sigma_{th} =1.1$  MeV for the external counters (mass $M=0.43$ ktons).

For SK we adopted the efficiency published for the Kamiokande-2 experiment \cite{Hirata:1988ad}.
\\

\noindent
4. The energy resolution $\Delta$, which appears in the resolution function (\ref{eq:resfun}).   We followed ref.  \cite{Bahcall:1997ha} for SK and SNO experiments, and refs. \cite{NIMA309,LVDprivcom} for LVD. The parameter $\Delta$  depends on energy as follows:
\be
{\Delta \over {\rm MeV}}=A {E_e \over {\rm MeV}} + B \sqrt{{E_e \over {\rm MeV}}}~.
\label{a2}
\ee
The values of the coefficients $A$ and $B$ are given in Table \ref{tab:tab1}.

\begin{table}[hbt]
\centering
\begin{tabular}{|c|c|c|c|c|c|c|}
\hline
detector  & $\delta$ &  $\alpha$ & material & mass (ktons) & A  & B
 \\
\hline

SK   & $36^\circ 21'$  & $137^\circ 18'$   & ${ \rm H_2 O}$  & 32  & 0  & 0.5 \\ 

\hline

SNO  & $46^\circ 30'$  & $-81^\circ 01'$   & ${ \rm H_2 O}$ & 1.4  &  0  &  0.35  \\
  &    &        & ${ \rm D_2 O}$  & 1.0  &          &         \\

\hline

LVD  & $42^\circ 25' $ & $13^\circ 31'$   & ${\rm(C H_2)_{10}}$  & 1.0  &  0.07 & 0.23 \\
\hline
\end{tabular}
\caption{Summary of the characteristics of the detectors we consider, in particular their locations on the Earth (in terms of northern latitude $\delta$ and eastern longitude $\alpha$), the fiducial masses and the coefficients $A$ and $B$ which appear in the expression of the energy resolution, eq. (\ref{a2}).}
\label{tab:tab1}
\end{table}

\section*{Appendix B. The regeneration factor: step-like and realistic Earth profile}
\label{appb}
The analytical expressions for the   regeneration factors (\ref{regf}) and (\ref{reganti}) can be obtained in the two layers approximation of the Earth density profile. In this approximation the mantle and the core of the Earth are considered as layers with constant densities. Therefore the neutrinos experience a constant matter potential along trajectories in the mantle and a step-like profile mantle-core-mantle along core crossing trajectories.
In what follows we summarize the analytical results obtained in the two-layers approximation, which correctly describe the general features of the Earth matter effects. The comparison with exact numerical calculations will be given at the end of  this appendix.
\\

\noindent
For $\nu_e$ propagating in the mantle only the regeneration factor, $f_{reg}$, eq. (\ref{regf}), has the form:
\be
f_{reg}=D(E,\theta,\theta_n,\rho_m) \sin^2 \left(\pi {d \over l_m}\right)~,
\label{b1}
\ee 
where $d=2 R_\oplus \cos\theta_n$ is the length of the trajectory in the Earth, $R_\oplus\simeq 6400$ Km is the radius of the Earth and $\rho_m$ the matter density in the mantle.
The depth $D$ of oscillations equals \cite{Smirnov:1994ku}:
\be
D=\sin 2\theta_m \sin(2\theta_m-2\theta)~,
\label{b2}
\ee
with $\theta_m$ and $l_m$ being the mixing angle and the oscillation length in matter and $\theta$ the mixing angle in vacuum.

Since the mixing is enhanced in matter, $\theta_m>\theta$, the depth $D$ is positive as well as the whole regeneration factor (\ref{b1}).
From eq. (\ref{b2}) one gets:
\be
D=x {\sin^2 2\theta  \over {(x-\cos 2\theta)^2+\sin^2 2\theta }  }~,
\label{b2b}
\ee
where 
$$x={2EV \over \Delta m^2}$$ 
and $V$ is the matter potential.
The expression (\ref{b2b}) vanishes in the limits of low ($x\ll 1$) and high ($x \gg 1$) energies.  It reaches the maximum
\be
D_{max}=\cos^2 \theta~,
\label{b3}
\ee  
at  $x=1$, which corresponds to the  energy:
\be
E_R={\Delta m^2 \over 2\sqrt{2} G_F n_e}~.
\label{b4}
\ee
Here $G_F$ is the Fermi constant and $n_e$ the electron number density in the medium. 

Thus, the depth of oscillations, $D$, has a resonant character with $E=E_R$ being the resonance condition. 
The width of the resonance is given by the interval $E_-\div E_+$ in which $D\geq D_{max}/2$.  One finds:
\be
{ E_{\pm} \over E_R}=2-\cos 2\theta \pm \sqrt{(1-\cos 2\theta)(3-\cos 2\theta)}~,
\label{b5}
\ee
which shows that the peak at $E\sim E_R$ is wide for LMA oscillation parameters and 
gets narrower as $\theta$ decreases. Notice that in the limit $\theta \rightarrow 0$ the maximal depth increases, $D_{max}\rightarrow 1$, but the oscillations disappear due to the decrease of the oscillation phase and the  vanishing of the resonance width (see eq. (\ref{b5})). 
\\

For antineutrinos, and trajectory in the mantle only, the regeneration factor ${\bar f}_{reg}$, eq. (\ref{reganti}), has the same form as in eq. (\ref{b1}), with the oscillation depth
\beq
{\bar D}&=&-\sin 2\theta_m \sin(2\theta_m-2\theta)~, \nonumber \\
&=& x {\sin^2 2\theta  \over {(x+\cos 2\theta)^2+\sin^2 2\theta } }~,
\label{b6}
\eeq 
where we defined $$x\equiv {2E|V|\over \Delta m^2}~.$$ 
Similarly to the case of $\nu_e$ the depth $\bar D$, and therefore the regeneration factor ${\bar f}_{reg}$, is positive, since $\theta_m<\theta$; moreover it has a similar resonant behaviour 
with maximum
\be
{\bar D}_{max}=\sin^2 \theta~,
\label{b7}
\ee
and the same resonance energy, eq. (\ref{b4}).
The borders $E_-,E_+$ of the resonance interval are given by eq. (\ref{b5}) with the replacement $\cos 2\theta \rightarrow - \cos 2\theta$. In  the limit $\theta \rightarrow 0$ the Earth effect disappears due to the vanishing of the oscillation depth (\ref{b7}).
For maximal mixing, $\cos 2\theta=0$, we get 
$$D={\bar D}={x  \over x^2+1}~.$$
\\

\begin{figure}[hbt]
\begin{center}
\epsfig{file=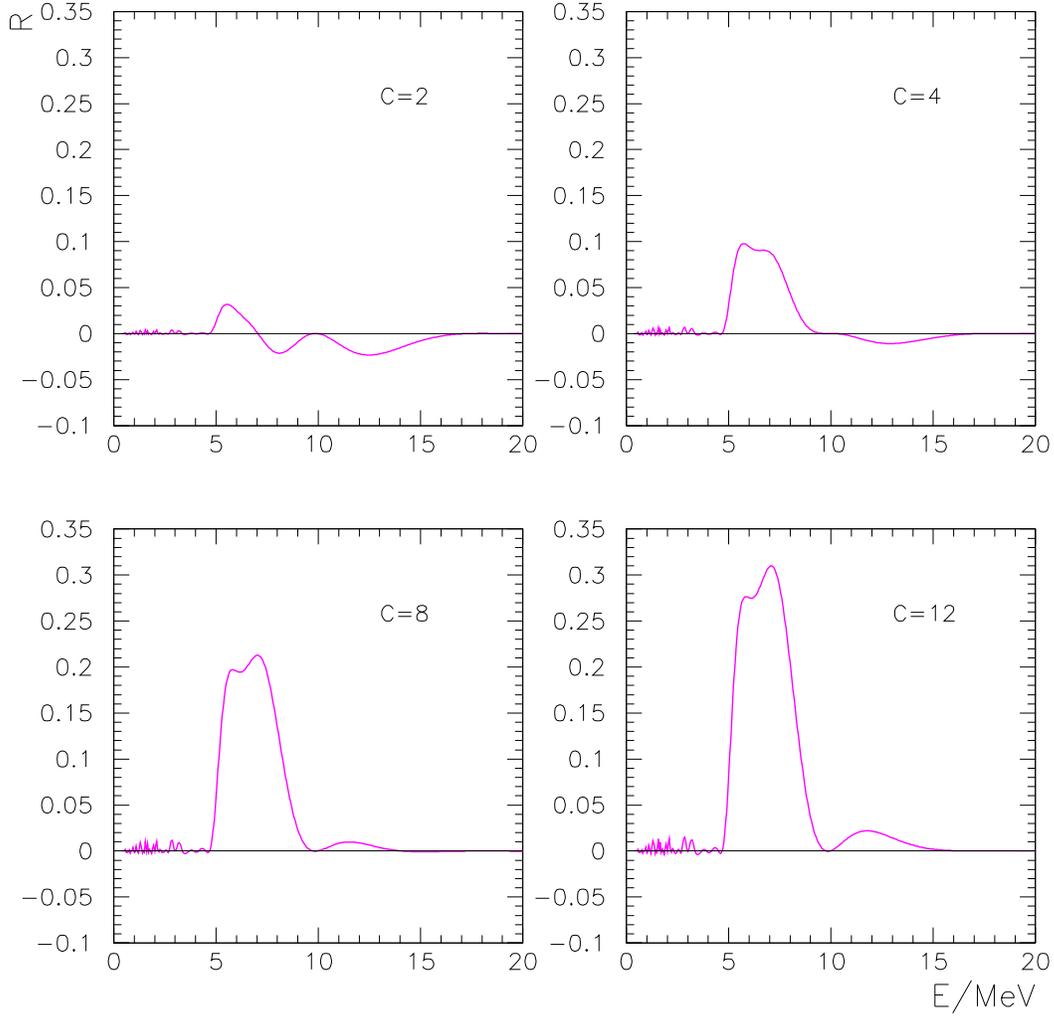, width=15truecm}
\end{center}
\caption{The relative Earth matter effect in $\nu_e$ channel, $R$, as function of the neutrino energy for SMA oscillation parameters and various values of the density profile factor  $C$. We have taken $\Delta m^2_{\odot}=6\cdot 10^{-6}~{\rm eV^2}$, $\sin^2 2\theta_{\odot}=5\cdot 10^{-3}$;  $T_e=3.5$ MeV, $T_x=8$ MeV; $P_H=1$ (or inverted hierarchy);
 $\theta_n=0^\circ$.   
} 
\label{fig:fig9} 
\end{figure}

\begin{figure}[hbt]
\begin{center}
\epsfig{file=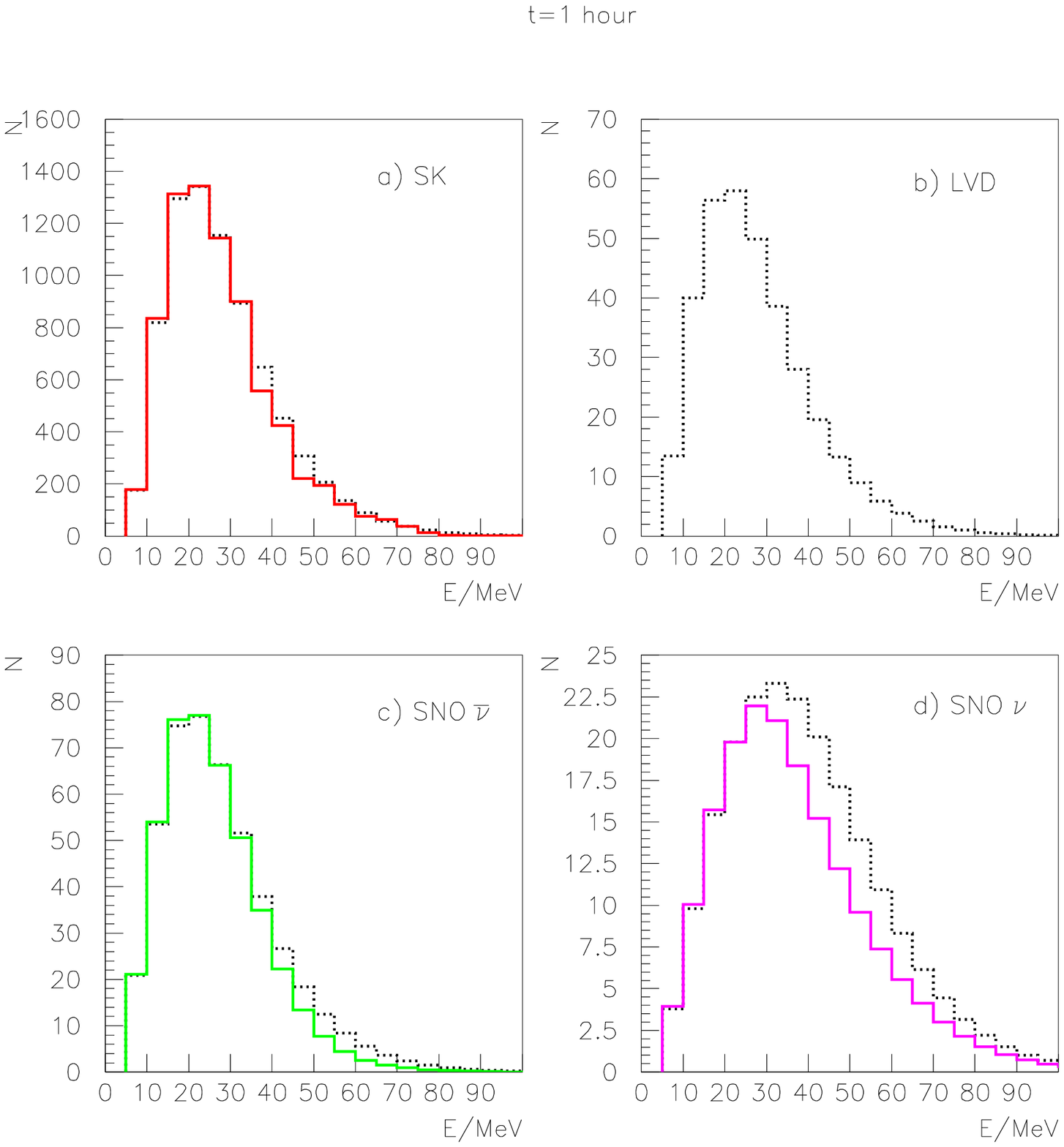, width=17truecm}
\end{center}
\caption{The energy spectra expected at SK, SNO and LVD with (solid lines) and without (dotted lines) Earth matter effect, for the same parameters as in figs. \ref{fig:fig5} and \ref{fig:fig1} and $t=1 $ hour of fig. \ref{fig:nadir} a).
A distance $D=10$ Kpc from the supernova and  binding energy $E_B=3\cdot 10^{53}$ ergs have been taken. In this specific configuration LVD is not shielded by the Earth, thus observing undistorted spectrum.
 The histogram c) refers to the sum of events from ${\bar \nu}_e +p\rightarrow e^+ + n$ and ${\bar \nu}_e + d\rightarrow e^+ +n +n$ scatterings, while the panel d) shows the events from ${\nu}_e + d\rightarrow e +p +p$. In a) and b) only the events from  ${\bar \nu}_e + p\rightarrow e^+ + n$ are shown. 
} 
\label{fig:histt1} 
\end{figure}

\begin{figure}[hbt]
\begin{center}
\epsfig{file=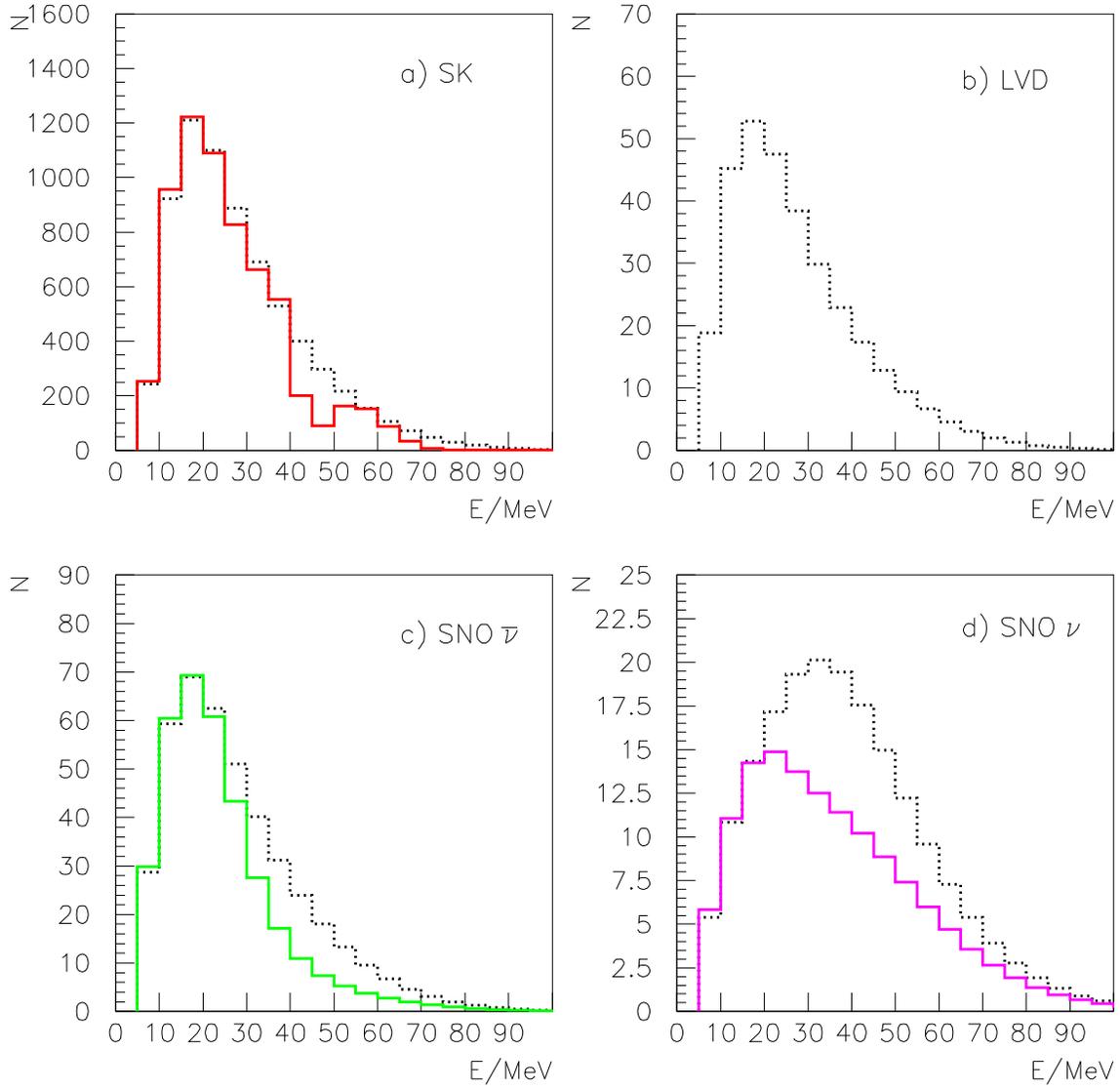, width=17truecm}
\end{center}
\caption{The same as fig. \ref{fig:histt1} for different values of some parameters: $\Delta m^2_{\odot}=3 \cdot 10^{-5}~{\rm eV^2}$, $\sin^2 2\theta_{\odot}=0.9$;  $T_e=3$  MeV, $T_{\bar e}=4$ MeV.
} 
\label{fig:histopt} 
\end{figure}

\begin{figure}[hbt]
\begin{center}
\epsfig{file=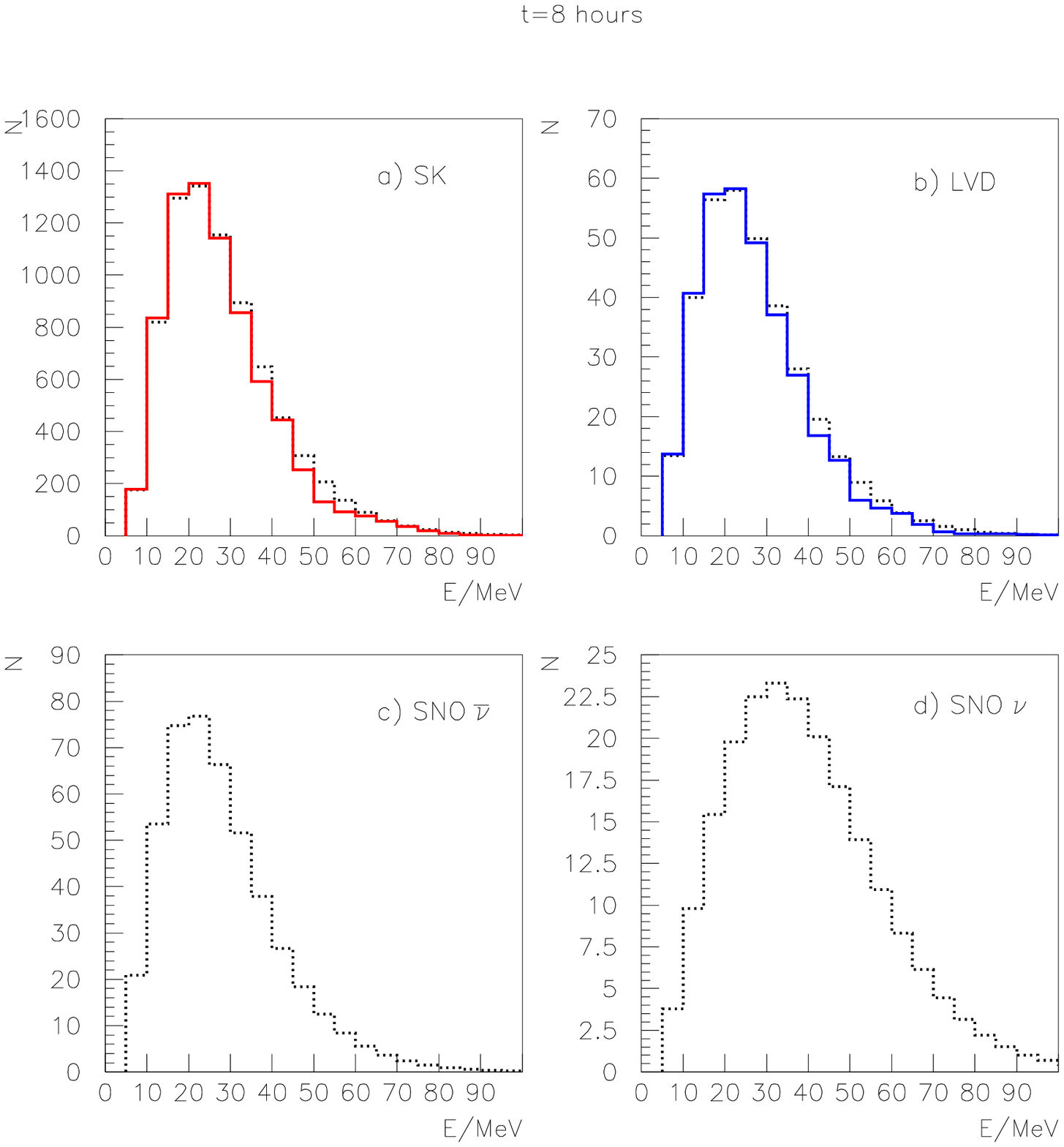, width=17truecm}
\end{center}
\caption{The same as fig. \ref{fig:histt1} for $t=8$  hours of  fig. \ref{fig:nadir} a). For this configuration SNO is unshielded by the Earth.
} 
\label{fig:histt8} 
\end{figure}

\begin{figure}[hbt]
\begin{center}
\epsfig{file=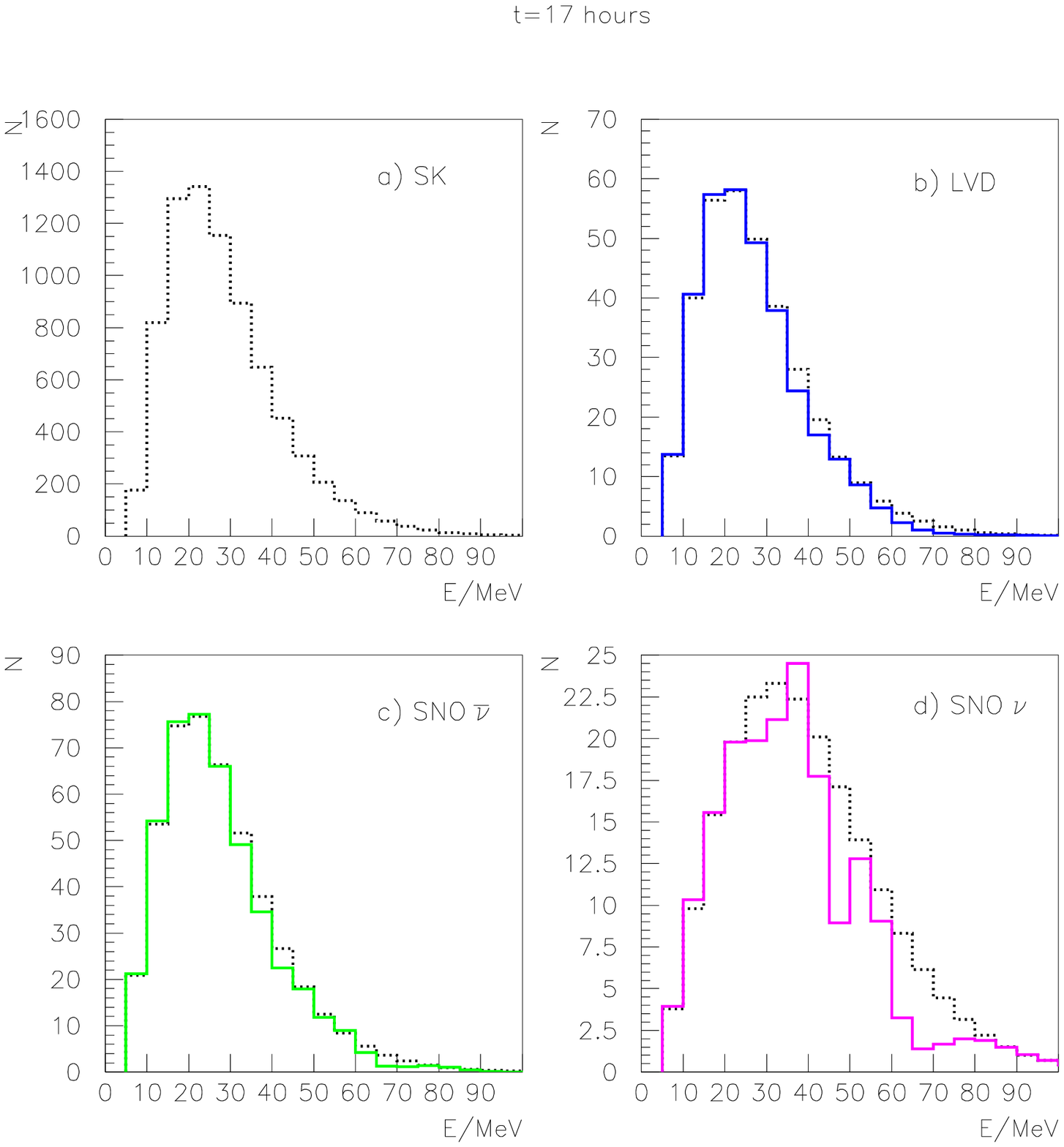, width=17truecm}
\end{center}
\caption{The same as fig. \ref{fig:histt1} for $t=17$  hours of  fig. \ref{fig:nadir} a).  For this configuration SK is unshielded by the Earth.
} 
\label{fig:histt17} 
\end{figure}

\begin{figure}[hbt]
\begin{center}
\epsfig{file=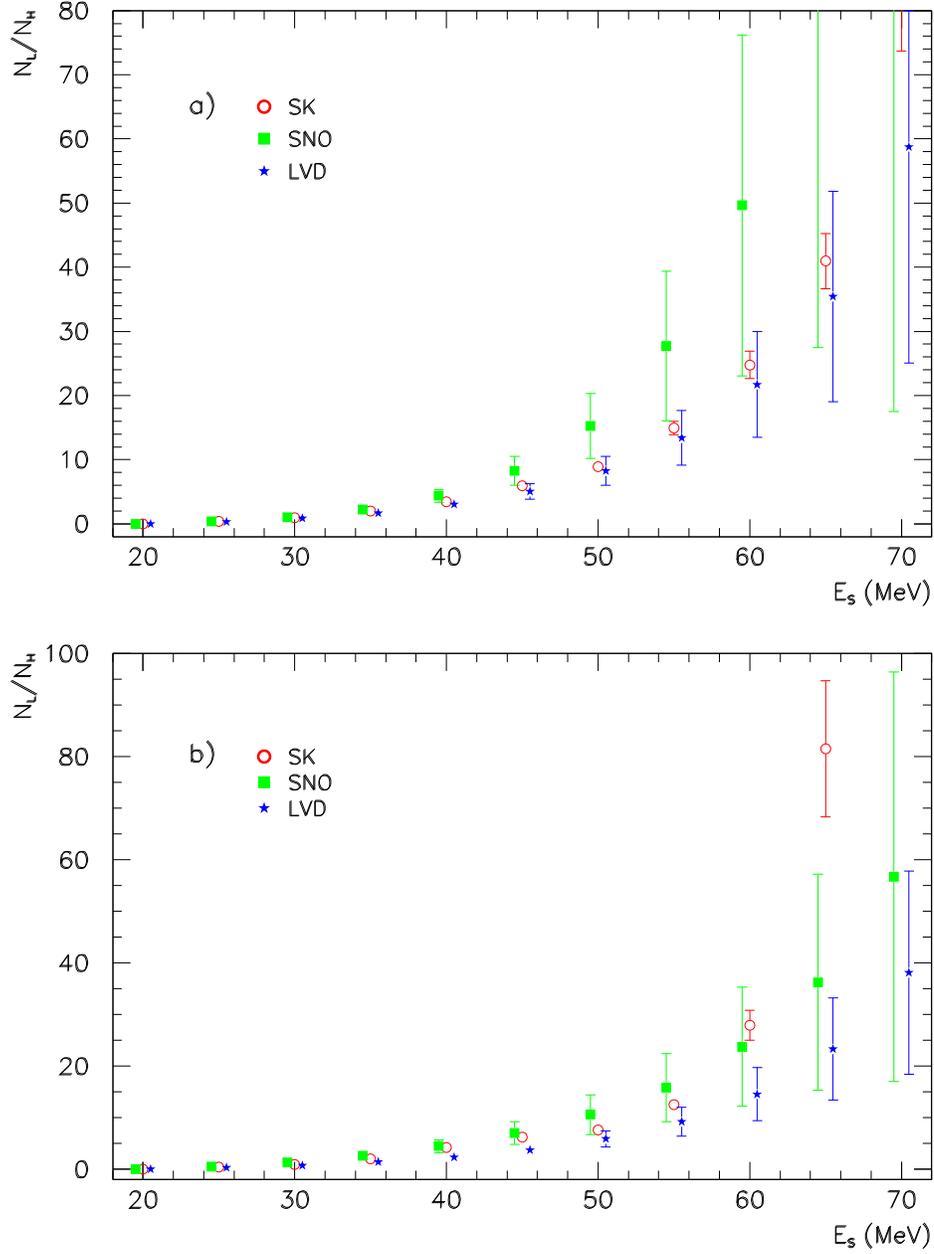, width=14truecm}
\end{center}
\caption{The ratio of the numbers of low and high-energy events from $\barnue +p\rightarrow e^+ + n$ reaction at SK, SNO and LVD, as a function of the separation energy $E_s$. The  bars represent $1~\sigma$ statistical errors.
The  panels a) and b) refer to the spectra shown in figs. \ref{fig:histt1} and  \ref{fig:histopt} respectively.
We have taken a minimum energy $E_{th}=20$ MeV for the calculation of the numbers of low-energy events. 
} 
\label{fig:ratios} 
\end{figure}

If the neutrino trajectory crosses both the mantle and the core the analytical treatment of the regeneration factors, ${\bar f}_{reg}$ and ${f}_{reg}$, is more complicated \cite{Akhmedov:1999ty}. The interplay of oscillations in the mantle and in the core determine irregular oscillations of the factors with energy. The depth of oscillations is larger in the region of the energy spectrum close to the resonance energies in the mantle and in the core; for SMA oscillation parameters parametric effects appear (see fig. \ref{fig:fig9}). In contrast to the propagation in the mantle only (i.e. in uniform medium), the regeneration factors have negative sign in some intervals of energy.\\

\begin{figure}[hbt]
\begin{center}
\epsfig{file=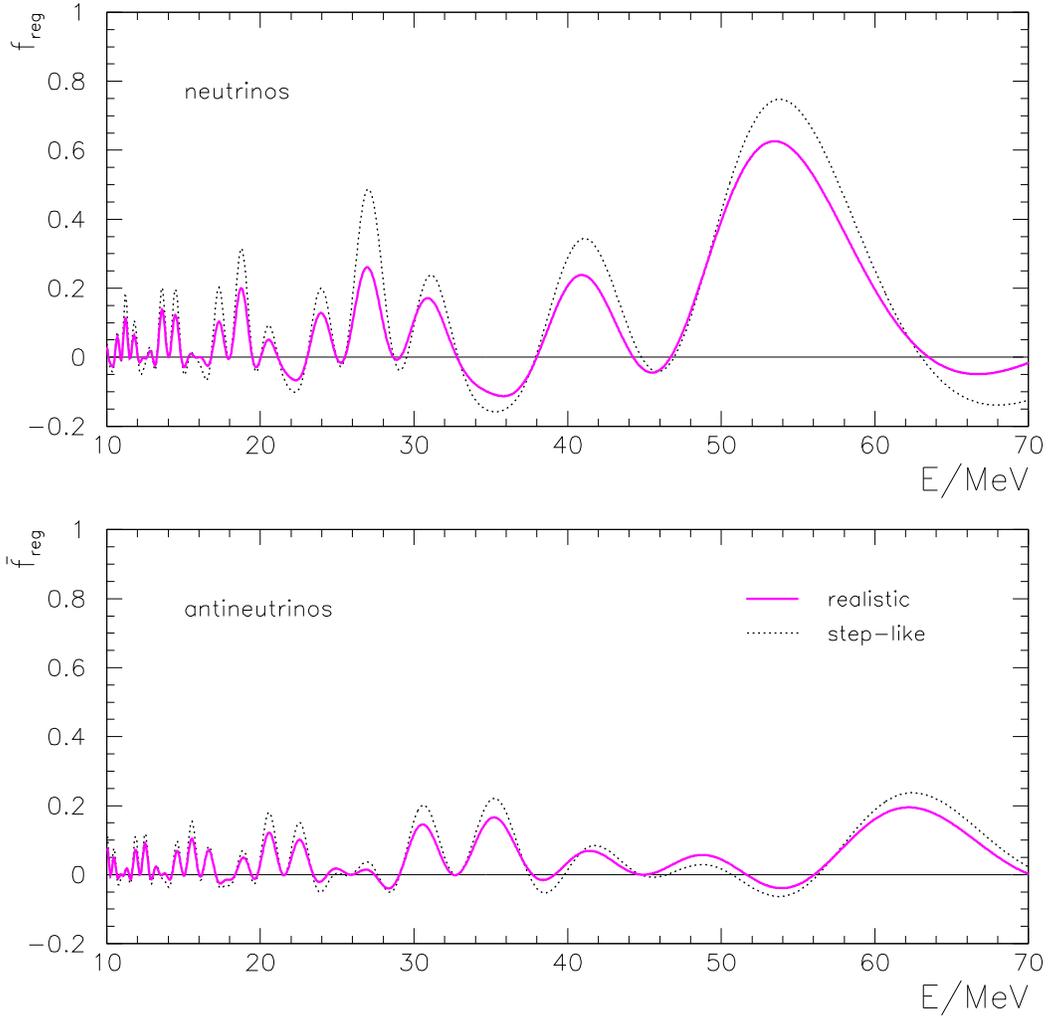, width=15truecm}
\end{center}
\caption{The regeneration factors for neutrinos, $f_{reg}$, and antineutrinos, ${\bar f}_{reg}$, calculated with step-like (two layers) and realistic profiles. We have taken $\Delta m^2=5\cdot 10^{-5}~{\rm eV^2}$, $\sin^2 2\theta=0.75$, $\theta_n=0^\circ$ and the densities $\rho_m=4.51$ and $\rho_c=11.95$ for the two-layers profile. 
} 
\label{fig:realistic} 
\end{figure}

Once a realistic density profile of the Earth is considered, the calculation of the regeneration factors requires a numerical treatment.
The results are presented in fig. \ref{fig:realistic} together with the analytical curves obtained with the two-layers approximation. 
The figure shows the factors ${\bar f}_{reg}$ and ${f}_{reg}$ as functions of the neutrino (antineutrino) energy for $\Delta m^2=5\cdot 10^{-5}~{\rm eV^2}$, $\sin^2 2\theta=0.75$ and  $\theta_n=0^\circ$.
We used the realistic profile in ref. \cite{PREM} and chose a step-like (two-layers) profile with densities $\rho_m=4.51$ and $\rho_c=11.95$, corresponding to the average densities of the profile of ref. \cite{PREM} in the mantle and in the core along the diameter of the Earth.

From the figure \ref{fig:realistic} it follows that the position of the oscillation maxima and minima on the energy axis are well reproduced by the step profile. 
This good approximation of the oscillation phase is ascribed to the choice of $\rho_m$ and $\rho_c$ to be the average densities of the two layers along the trajectory of the neutrinos.

In contrast, the  depth of oscillations given by the numerical calculation deviates significantly, up to $\sim 50\%$, from the result of the analytical (two-layers) approximation. As a general tendency, the depth of oscillations in the realistic density profile appears smaller with respect to the case of the two-layers profile.
If the density jumps along the trajectory are not very large  (e.g. for trajectories in the mantle only) this feature can be interpreted, qualitatively, by considering the density as smoothly varying along the path of the neutrinos. In this case the neutrino conversion occurs adiabatically and the depth of oscillations is determined by the matter density at the surface of the Earth. Since the surface density is smaller than the average density along the trajectory and the latter in turn is smaller than the resonance density $\rho_R$ in the relevant range of energies, a smaller depth of oscillations is expected. 

A better approximation of the numerical results can be obtained by using the average density in the determination of the oscillation phase and the surface density in the determination of the depth of oscillations.

For SMA parameters the adiabaticity is broken and the two layers model gives a good  approximation of the numerical results.

\bibliography{draft}

\clearpage

\end{document}